\documentclass[
reprint, 
superscriptaddress, nofootinbib, amsmath, amssymb, aps, prx,
floatfix,
]{revtex4-2}

\usepackage{siunitx}
\usepackage[normalem]{ulem}
\usepackage[inline]{enumitem}
\usepackage{mathtools}
\usepackage{graphicx}% Include figure files
\usepackage{dcolumn}% Align table columns on decimal point
\usepackage{bm}% bold math
\usepackage[dvipsnames]{xcolor}
\usepackage[linkcolor=Blue,citecolor=Blue,urlcolor=Blue,colorlinks=true]{hyperref}

\DeclareMathOperator{\tr}{tr}
\usepackage{amsmath,amssymb}

\def\vec{\boldsymbol}
\def\mat{\boldsymbol}

\DeclareMathOperator{\MSD}{MSD}

\begin{document}

\title{%
Externally driven condensates show translation-induced polarization, directed coalescence, and anomalous diffusion in viscoelastic media
}%

\author{Andriy Goychuk}
\email{andriy@goychuk.me}
\affiliation{Institute for Medical Engineering and Science, Massachusetts Institute of Technology, Cambridge, MA 02139, United States}
\affiliation{Department of Systems Immunology, Helmholtz Centre for Infection Research, 38124 Braunschweig, Germany}
\affiliation{Institute for Biochemistry, Biotechnology and Bioinformatics, Technische Universität Braunschweig, Braunschweig, Germany}
\affiliation{Lower Saxony Center for Artificial Intelligence and Causal Methods in Medicine (CAIMed), Hannover, Germany}

% Please add a significance statement to explain the relevance of your work
%Living cells often form small droplets when biomolecules in the cytoplasm and nucleoplasm demix.
%Rearranging these droplets via chemical stimuli or mechanical forces is an attractive route to control cell organization. 
%However, it hinges on the assumption that local perturbations show local effects.
%In contrast, we show analytically that dragging a droplet universally induces long-range, dipole-like forces that act on other droplets and drive directed coalescence.
%Because this finding distinguishes liquid droplets from solid colloids, we revisit the fundamental phenomenon of Brownian motion.
%Active mechanical stresses and fluctuations lead to anomalous, time scale-dependent diffusion with a non-Stokesian relation to condensate size due to correlated fluid flows.
%These results constitute a step towards understanding the nonequilibrium motion of condensates in cells.

% Please include corresponding author, author contribution and author declaration information
%\authorcontributions{A.G. designed research, carried out analytical calculations and related numerical implementation, analyzed the results, and wrote the manuscript.}

%\authordeclaration{There are no competing interests to declare.}

% At least three keywords are required at submission. Please provide three to five keywords, separated by the pipe symbol.

%\author{Andriy Goychuk}
%\email{andriy@goychuk.me}
%\affiliation{Institute for Medical Engineering and Science, Massachusetts Institute of Technology, Cambridge, MA 02139, United States}%

\date{\today}% It is always \today, today,

\begin{abstract}
Phase separation into compositionally and physically distinct domains is ubiquitous in (non)living matter ranging from alloys and emulsions to biomolecular condensates in cells.
The organization of these domains can be controlled, for example, by nonequilibrium chemical reactions, external fields, or mechanical stresses.
In this context, stationary states can emerge from effective long-range interactions resembling the electrostatics of charges.
As shown here, externally controlled dynamic states, such as condensate motion, lead to an effective polarization and dipolar force fields even for microscopically nonpolarizable matter.
The dipole-dipole interactions resulting from this \emph{translation-induced polarization} cause directed coalescence of domains.
This coarsening mechanism complements Ostwald ripening and coalescence due to Brownian motion or Marangoni flows, and has implications for controlling domains by electric fields or concentration gradients.
Interestingly, the chemical potential gradients around a domain that nucleates material are exactly opposite to the hydrodynamic pressure gradients around an impermeable colloid that pushes the fluid, suggesting a competition between phase separation and hydrodynamics.
In addition to chemical control, the motion of domains can also be driven by mechanical stresses.
An example is the cell interior, where mechanical stresses are actively generated by molecular motors and opposed by passive viscoelastic stresses in the cytoplasm and nucleoplasm.
The resulting fluid flows lead to Brownian motion with a suppressed or enhanced size scaling which modifies collision-coalescence.
For active stresses with a long correlation time, the domains show superdiffusion on intermediate time scales.
Together, these findings shed new light on the dynamics of domains in viscoelastic media and conserved order parameters in general.
\end{abstract}

\maketitle
%\thispagestyle{firststyle}
%\ifthenelse{\boolean{shortarticle}}{\ifthenelse{\boolean{singlecolumn}}{\abscontentformatted}{\abscontent}}{}

%\firstpage[6]{3}

%\tableofcontents

\section{Introduction}
Phase separation is a recurrent phenomenon in various disciplines.
Its theoretical description was originally developed in the context of polymer melts~\cite{huggins_1941, flory_1941, flory_1942} and alloys~\cite{cahn_hilliard_1958, cahn_1961}.
Since then, the applications and implications of phase separation have spanned soft matter, materials science, and (chemical) engineering of, for example,  batteries and reactors~\cite{bazant_2013, joshi_2015}.
More recently, the formation of biomolecular condensates via phase separation has been recognized as an organizing principle in cell biology~\cite{brangwynne_2009, hyman_2014, banani_2017, lyon_2020, alberti_2019, shin_2017, choi_2020}.
The underlying mechanism is that (bio)molecules can spatially self-organize by demixing and sorting  according to their relative miscibility.
In the cell nucleus, for example, biomolecular condensates have been implicated in vital processes such as gene transcription~\cite{hnisz_2017, sabari_2018, shrinivas_2019, henninger_2021, hirose_2023}, splicing~\cite{faber_2022}, and assembly of ribosomal subunits~\cite{lafontaine_2021, banani_2024}.

The central feature of phase separation in an equilibrium setting is the gradual minimization of a free energy dominated by the interfacial tension between the different phases.
Consequently, different condensates undergo coarsening until only a single, spherical interface with minimal contact area between the different phases is left.
The kinetics of coarsening can be driven by several mechanisms. 
In Ostwald ripening, surface tension and the resulting Laplace pressure cause large condensates to gradually siphon material from small condensates via currents in the dilute phase~\cite{voorhees_1992}.
In addition, condensates that behave like liquids will also coalesce when they encounter each other due to Marangoni flows or Brownian motion~\cite{siggia_1979, shimizu_2015}.
Coalescence by Brownian motion requires hydrodynamic fluctuations, which many theoretical studies neglect.

To functionalize phase-separated domains and condensates in the context of engineering and biology, their spatial positioning and coarsening must be controlled.
However, it is not clear how forcing one condensate to move will affect other, nearby condensates.
This is a classical problem in the context of rigid colloidal particles embedded in a viscous fluid~\cite{happel_brenner_1983, driscoll_2019}, but there is no theoretical justification to treat liquid-like condensates as rigid colloids.
Analogously, there is well-established literature on the Brownian motion of colloidal particles in complex media~\cite{frey_kroy_2005, squires_2010}, yet the motion of condensates embedded in viscoelastic media~\cite{tanaka_2000} has barely been explored.

In the context of cell biology, the functions of some condensates involve interactions with the cytoskeleton~\cite{wiegand_2020, mohapatra_2023, graham_2024} or chromatin~\cite{quail_2021, strom_2024}.
This coupling renders condensates sensitive to viscoelastic stresses.
Conversely,  the cell cytoskeleton generates active stresses~\cite{fletcher_2010}, chromatin shows correlated flows~\cite{zidovska_2013, bruinsma_2014, shaban_2018}, and the viscoelastic medium can be actively stirred.
Mechanical forces can drive ripening or arrest coarsening~\cite{style_2018, rosowski_2020, yuan_2021, qiang_2024, yu_2023, winter_2024}, and stirring can generally suppress phase separation~\cite{taylor_1934, stone_1994} as observed in active nematic fluids~\cite{caballero_2022, adkins_2022}.

Yet, active forces are also known to accelerate the diffusion of rigid particles~\cite{brangwynne_2009_active_diffusion}.
Extrapolating these insights suggests that active forces should facilitate the coalescence of biomolecular condensates, as was recently observed for speckles in the cell nucleus~\cite{aljord_2022}.
However, beyond this qualitative expectation, the Brownian motion of condensates in active viscoelastic media is not well understood.
% WHAT DOES THIS MANUSCRIPT CONTRIBUTE SPECIFICALLY
The main goal of the present work is to develop a theory for the motion of condensates and phase-separated domains in a nonequilibrium setting.
To that end, using an exact analytical framework, I will address two connected paradigmatic questions as outlined below.

First, how do pairs of condensates interact when they move due to an external potential gradient, such as gravitation, an electrostatic potential, or a concentration gradient?
In analogy to viscous fluids and electrostatics, where incompressibility and charge conservation lead to nonlocal disturbances and interaction forces~\cite{happel_brenner_1983}, I will show that mass conservation leads to similar effects in phase-separating solutions~\cite{bray_1994}.
Even in media that cannot be polarized microscopically, the motion of condensates induces chemical potential gradients (\emph{translation-induced polarization}) that are reminiscent of the electrical field around an electrostatic dipole.

Surprisingly,  the translation-induced chemical potential gradients are opposite to the pressure gradients generated by colloidal particles moving through a viscous fluid. 
The intuition is that, contrary to the mechanics of impermeable domain walls which push fluid, phase-separated domains nucleate material through a gain in enthalpy at the phase boundaries.
This suggests a generic competition between phase separation and hydrodynamics.
Moreover, the translation-induced chemical potential gradients cause small condensates to drift towards the front of nearby larger condensates, leading to directed coalescence.
This prediction complements the ripening of condensates under external fields~\cite{weber_2017, Bressloff_2020} and can be tested experimentally~\cite{jambonpuillet_2023, doan_2024} by measuring the time-dependent distribution of condensate radii.
Moreover, it implies that coherent flows of chromatin on the length scale of microns~\cite{zidovska_2013} can, in addition to bringing together condensates over larger distances~\cite{aljord_2022}, drive the deterministic coalescence of differently sized condensates on a sub-micron scale.

Second, how will active fluctuations in a viscoelastic medium such as the nucleoplasm drive the Brownian motion of individual condensates?
In the context of colloids,  this fundamental question has led to the Stokes-Einstein relation~\cite{einstein_1905, frey_kroy_2005}, which couples fluctuations and dissipation, and which has been foundational for the field of soft and biological matter.
In the context of phase-separated domains, however, studies considering Brownian motion are more scarce~\cite{schmitt_stark_2016, cates_review_2018, wilken_2023, lee_2023, zhang_2024_brownian}.
To close this gap, I will derive a principled theory of the time-dependent mean squared center-of-mass displacement of condensates and spatially extended fields in fluctuating media with memory.

Finally, I will apply the theory to study the Brownian motion of condensates in the cell cytoplasm and nucleoplasm, which are idealized as viscoelastic Maxwell fluids driven by active stress fluctuations.
Active stress fluctuations violate the fluctuation-dissipation theorem and, by extension, the Stokes-Einstein relation. 
This alters the kinetics of droplet coalescence by suppressing or amplifying the scaling of the center-of-mass diffusion coefficient with condensate radius.
If the correlation length of the fluid flows (stirring length) is larger than the condensate radius, then faster Brownian motion due to viscoelastic stirring will only decrease the time until two condensates are separated by one stirring length.
On distances shorter than the stirring length, condensate motion will become hydrodynamically locked by the coherent flows, thus decreasing the fluctuations in the mean squared separation between condensates.
Then, directed coalescence by translation-induced polarization will become the dominant mechanism.

Thus, the two paradigmatic questions addressed in this work shine light on two sides of the same coin, by discussing how Brownian motion can first move condensates closer together and how then, on short length scales, capillary forces can guide the coalescence process to completion.

\section{Phase separation under turnover and external fields}
\subsection{Mass conservation}
Biomolecular condensates refer to regions with increased concentration $c(\vec{x},t)$ of macromolecules such as proteins or nucleic acids~\cite{shin_2017}.
They form due to pairwise interactions (enthalpy), which tend to cluster molecules, competing against entropy which homogenizes the system.
This competition is embodied in the local free energy density $f(c)$, which can be derived from Flory-Huggins theory~\cite{huggins_1941, flory_1941, flory_1942} or from symmetry arguments within a Ginzburg-Landau expansion~\cite{cahn_hilliard_1958, cahn_1961}.
In the simplest case, the free energy density has two minima which correspond to the high-concentration region and the low-concentration background.
The total free energy of a two-component solution, including the energetic cost of phase boundaries with stiffness parameter $\kappa$, is given by
%
% IN THE END THIS WILL BE IRRELEVANT, SO JUST MAKE SURE IT IS CORRECT AND WE DO NOT DOUBLE-DIP SYMBOLS
\begin{equation}
    \mathcal{F}[c] = \int d^d\! \vec{x} \, \left[ f(c) + \frac{1}{2} \kappa \, (\vec{\nabla}c)^2 \right] \, .
\end{equation}
The chemical potential, $\mu(\vec{x}) \coloneqq \delta\mathcal{F}[c]/\delta c(\vec{x})$, corresponds to the energetic cost of adding one macromolecule at coordinate $\vec{x}$.
Phase separation is a consequence of fluxes that shuffle molecules (redistribute mass) along the gradients in the chemical potential, until the system has settled in thermodynamic equilibrium~\cite{balian_2006, degroot_2013}.

To control this process, one can apply an external potential $\Psi(\vec{x},t)$.
This potential could correspond to, for example, an electrostatic or gravitational field or molecular interactions with a gradient in the concentration of another molecule.
Moreover, by allowing the potential $\Psi$ to fluctuate, one can model stochastic forces and disorder.
In summary, the material fluxes of the phase-separating molecules are given by
\begin{equation}
    \vec{J} = -M\vec\nabla\mu - M\vec\nabla\Psi + \vec{j}
\end{equation}
with $M$ the collective mobility and $\vec{j}$ the remainder of the fluxes that cannot be written as the streamlines of some potential.
Such nonpotential fluxes could correspond to, for example, advection $\vec{j} = \vec{v} c$ by fluid flow with velocity $\vec{v}$ as discussed later.

\subsection{Chemical turnover}
Biologically relevant macromolecules are generally turned over on some time scale $k^{-1}$ in the cell.
To model these chemical reactions,  I will consider exchange with an external reservoir maintained at a chemical potential $\mu^\star$.
To enter or leave the reservoir, biomolecules need to overcome a chemical potential barrier and reach a transition state corresponding to some chemical potential $\bar{\mu}$.
Approximating the barrier-crossing process with Arrhenius kinetics, biomolecules will be degraded (enter the reservoir) at rate $k_- e^{-\beta (\bar{\mu} - \mu)}$, with the standard notation $\beta = 1/(k_B T)$ and $\bar{\mu} - \mu$ the energetic cost of a molecule reaching the transition state.
Here, for simplicity, $\bar{\mu}$ has been chosen such that it absorbs the external potential $\Psi$.
Analogously, biomolecules will be produced with recovery rate $k_+ e^{-\beta (\bar{\mu} - \mu^\star)}$.
Thermodynamic consistency requires that the ratio between the production rate and the degradation rate equals the Boltzmann weight corresponding to the energy difference between the domain and the reservoir, which constrains the rate constants $k_+ = k_- \equiv k$~\cite{zwicker_2022, zwicker_2025}.

In summary, the dynamics of the concentration profile $c(\vec{x},t)$ of biological macromolecules is given by
\begin{multline}
\label{eq:continuity_nonlinear}
    \partial_t c = \vec\nabla{\cdot}\left[M\vec\nabla\mu + M\vec\nabla\Psi - \vec{j} \right] \\
    + k e^{-\beta (\bar{\mu} - \mu^\star)} - k e^{-\beta (\bar{\mu} - \mu)} \, .
\end{multline}
This general formalism can also be adapted beyond the dynamics of condensates to conserved Ising models, or active matter, to name a few examples.
At first glance, it seems that the dynamics of Eq.~\eqref{eq:continuity_nonlinear} is highly nonlinear and should depend on the microscopic details of the molecular interactions encoded in the chemical potential $\mu$.
However, note that the analysis started with the premise of phase separation, that is, a macroscopic phenomenon.
In the spirit of thermodynamics, where macroscopic phenomenology allows making general statements independent of microscopic details, I will therefore focus on the consequences of phase separation, irrespective of how it is achieved.
To that end, following Ref.~\cite{goychuk_2024_self_consistent} which semi-analytically studied condensate self-propulsion in the absence of hydrodynamics, in the following Eq.~\eqref{eq:continuity_nonlinear} will be interpreted as a control problem and the chemical potential as an unknown Lagrange multiplier.

\section{Condensate motion induces dipole forces}
\subsection{Effective Debye-Hückel theory}
It is well known that the chemical potential in phase separation problems satisfies a Poisson equation and thus mediates nonlocal interactions resembling electrostatics~\cite{bray_1994}.
This manifests as circulatory currents that accelerate the motion of individual condensates~\cite{goychuk_2024_self_consistent}, and as nonlocal friction during the motion and deformations of single droplets under thermophoresis~\cite{romano_2025preprint}.
As shown in the following, these nonlocal capillary interactions predict how pairs of condensates interact under external gradients and differ from colloids interacting hydrodynamically in viscous fluids.

From the perspective of electrostatics, Eq.~\eqref{eq:continuity_nonlinear} is reminiscent of the Poisson-Boltzmann equation.
For small $\beta$, or chemical reactions with small potential barriers, one can use a linear approximation in analogy to Debye-Hückel theory~\cite{debye_huckel_1923},
\begin{multline}
\label{eq:continuity}
    \vec\nabla{\cdot}\left[M\vec\nabla\mu\right] 
    - k \beta \mu 
    = 
    \partial_t c
    + \vec\nabla{\cdot}\left[\vec{j} - M\vec\nabla\Psi \right] 
    - k \beta \mu^\star
    \, ,
\end{multline}
where the left-hand side collects all terms involving the chemical potential.
The characteristic, diffusive screening length is given by $\sqrt{M / (k\beta)}$.
Substituting the mobility $M = k_B T / \zeta = 1/(\zeta \beta)$ with microscopic friction coefficient $\zeta$, reveals that the diffusive screening length diverges as $\beta^{-1} (\zeta k)^{-1/2}$ in the limit of high temperature, or small friction, or slow chemical turnover.
This sets an upper limit for the range of capillary forces.
Since the present work focuses on capillary interactions, the following analysis will focus on the limit of an infinitely large screening length.

\subsection{Heuristic analogy to electrostatics and fluid mechanics}
To develop an intuition for the chemical potential gradients $\vec{\nabla}\mu$ induced by condensate motion, consider a prescribed perturbation of the concentration field, $\partial_t c(\vec{x},t) \coloneqq \delta c(\vec{x}) \delta(t)$.
Mass exchange via a local addition (production) or removal (degradation) of material enters as a source or sink in the Poisson equation, Eq.~\eqref{eq:continuity}, and thus acts analogously to electric monopoles inducing an electrostatic potential~\cite{kumar_2023, zwicker_2025}.
In the case of a moving phase boundary, mass is removed from one side and added to the opposing side of the interface, analogous to a local electric dipole with source $\partial_t c(\vec{x},t) \coloneqq -\vec{\nabla}\cdot [\vec{v}(t) c(\vec{x},t)]$.
This suggests that translation will induce dipole force fields, whereas shape changes will lead to higher-order multipole fields, even in matter that is not polarizable microscopically.

One can also construct a similar analogy to fluid mechanics, where dipole force fields are induced by the motion of colloidal particles (spherical inclusions) through a viscous fluid.
To make this comparison more salient, the following analysis will consider condensates with spherical shape and radius $R$.
This is a valid approximation for high surface tension, but neglects multipole interactions due to capillary waves.
Assuming stable condensates, their domains $\mathcal{D}$ will change mainly due to translation and will be approximately conserved.
The role of the chemical potential is to maintain phase separation with a high concentration $c(\vec{x},t) = c_- + \Delta c$ inside the condensate, $\vec{x} \in \mathcal{D}$, and a low concentration $c(\vec{x},t) = c_-$ outside.
In that sense, the chemical potential can be interpreted as a Lagrange multiplier preserving the shape of the condensate boundary during translation, similar to how the pressure field enforces fluid incompressibility.

Based on these analogies to electrostatics and fluid mechanics, one expects that perturbations in the condensate interface propagate nonlocally and induce long-range forces.

\subsection{Translation-induced polarization}
Equipped with first intuitive predictions, I will now derive the chemical potential profile $\mu$ corresponding to Eq.~\eqref{eq:continuity} with some prescribed dynamics $\partial_t c$.
Condensates moving with velocity $\vec{v}$ are naturally described by the traveling wave ansatz, $c(\vec{x},t) = c(\vec{z})$ with the Galilean coordinate transformation $\vec{z} \coloneqq \vec{x}-\vec{v}t$.
In the following, $\vec\nabla \equiv \vec\nabla_{\vec{z}}$ refers to the spatial gradient and $\mathcal{D}$ to the condensate domain in the comoving frame whereas $\mathcal{D}(t)$ is the condensate domain in the laboratory frame.
For simplicity, as mentioned above, I will consider the limit of negligible material turnover, $k \to 0$.
Comparing Eq.~\eqref{eq:continuity} to a Poisson equation for the chemical potential~\cite{bray_1994}, with the fundamental solution $\mathcal{L}(\vec{z})$ satisfying $\vec\nabla^2\mathcal{L}(\vec{z}) = - \delta(\vec{z})$, reveals
\begin{multline}
\label{eq:chemical_potential}
    \mu(\vec{z}) = - \Psi(\vec{z}) 
    + \frac{\Delta c}{M} \, \vec{v} \cdot \int_{\mathcal{D}} d^d\vec{z}' \, \vec\nabla \mathcal{L}(\vec{z} - \vec{z}') \\
    - \frac{1}{M}\int d^d\vec{z}' \, \vec{j}(\vec{z}') \cdot \vec\nabla \mathcal{L}(\vec{z} - \vec{z}') \, .
\end{multline}
Hence, the gradient of the chemical potential in the comoving frame of the condensate is given by
\begin{multline}
\label{eq:long_range_force_comoving}
    \vec\nabla\mu(\vec{z}) = 
    - \vec\nabla\Psi(\vec{z}) 
    + \frac{\Delta c}{M} \, \mathcal{B}(\vec{z}) \cdot \vec{v} \\
    - \frac{1}{M} \, \int d^d\vec{z}' \, \left[
    \vec\nabla \otimes \vec\nabla \mathcal{L}(\vec{z} - \vec{z}')
    \right] \cdot \vec{j}(\vec{z}')
    \, ,
\end{multline}
with $\mathcal{B}(\vec{z}) \equiv \mathcal{B}(\vec{z} | \mathcal{D})$ the coupling tensor at coordinate $\vec{z}$ relative to the centroid of the condensate, 
\begin{equation}
\label{eq:tensor_longranged_force_general}
    \mathcal{B}(\vec{z} | \mathcal{D}) \coloneqq \vec\nabla \otimes \vec\nabla \int_{\mathcal{D}} d^d\vec{z}' \, \mathcal{L}(\vec{z} - \vec{z}') \, .
\end{equation}
The above integral can be evaluated for spherical condensates with radius $R$ and is identical to the Green's tensor for an electrostatic dipole field:
\begin{equation}
\label{eq:tensor_longranged_force}
    \mathcal{B}(\vec{z}) 
    = - \begin{dcases}
    \frac{\mat{I}}{d} \,, & |\vec{z}| \leq R \, , \\
    \left[\frac{\mat{I}}{d} - \hat{\vec{z}}\otimes\hat{\vec{z}}\right] \left(\frac{R}{|\vec{z}|} \right)^{d} \,, & |\vec{z}| > R \, .
    \end{dcases}
\end{equation}
Here, I have defined $\hat{\vec{z}} \coloneqq \vec{z} / |\vec{z}|$.
Transforming Eq.~\eqref{eq:long_range_force_comoving} back into the laboratory frame, with shorthand notation $\vec{r} \equiv \vec{r}(t)$ for the centroid of the condensate, shows that condensate motion with velocity $\vec{v}(t) = \partial_t \vec{r}(t)$ induces an effective long-range force (Fig.~\ref{fig::motion_induced_phenomena}A),
\begin{equation}
\label{eq:long_range_force}
    -\vec\nabla\mu(\vec{x},t) = 
    \vec\nabla\Psi(\vec{x},t) 
    - \frac{\Delta c}{M} \, \mathcal{B}(\vec{x} - \vec{r}) \cdot \vec{v}(t) + \mathcal{O}[\vec{j}] 
    \, ,
\end{equation}
in agreement with the heuristic reasoning.
%where $\mathcal{O}[\vec{j}]$ is a functional of order $\vec{j}$.
In summary, the motion of condensates and phase-separated domains, regardless of the underlying driving force, universally induces dipole force fields with dipole strength $v\Delta c R^d / M$.
The strength of the induced dipole is directly proportional to the controlled drift velocity, the number of particles enriched in the condensate, and the drag that these particles experience.
In the following, I refer to this phenomenon as \emph{translation-induced polarization} as opposed to motion that arises from pre-existing polarity.

\subsection{Competition of translation-induced polarization with hydrodynamics}
\begin{figure}[t]
    \centering
	\includegraphics{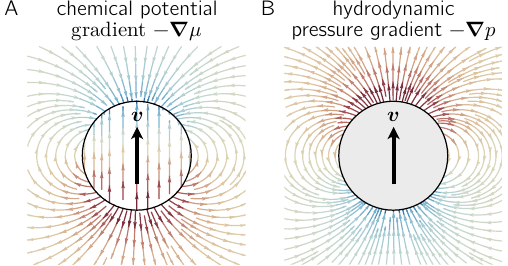}
	\caption{%
    \textbf{Comparison between a condensate and a colloidal particle} moving with velocity $\vec{v}$ (black arrows).
    A) Translational motion of domains of conserved order parameters (outlined by black line) induces long-ranged forces that are reminiscent of electrostatic dipoles.
    Streamlines indicate the gradient of the chemical potential (color code; red corresponds to high values and blue to low values).
    B) Impermeable boundaries push fluid when they move. 
    Therefore, the pressure gradients around moving colloidal particles in a Stokes fluid~\cite{landau_lifshitz_1987} are opposite to the chemical potential gradients around moving condensates.
    This suggests a competition between phase separation and fluid flow.
    Streamlines indicate the gradient of the pressure (color code; red corresponds to high values and blue to low values).
	}
	\label{fig::motion_induced_phenomena}
\end{figure}

This section will clarify important differences of condensate dynamics compared to the dynamics of colloids in viscous fluids, as revealed by the translation-induced chemical potential, Eq.~\eqref{eq:long_range_force}.
In contrast to active matter and hydrodynamic theories, where polarity refers to stresslet tensors generating force-free flow, here the chemical potential is a scalar thermodynamic field entering conserved dynamics.
The closest analog is a rigid sphere being dragged through fluid by an external force.
Moreover, because phase-separated interfaces nucleate material, whereas impermeable walls push the fluid, the chemical potential gradient is exactly opposite to the pressure gradient generated by a solid spherical particle with radius $a$ (Fig.~\ref{fig::motion_induced_phenomena}B),
\begin{multline}
    -\vec\nabla p = -\frac{3}{2} \eta a \vec\nabla \left( \frac{\vec{v} \cdot \vec{z}}{|\vec{z}|^3} \right) \\
    = -\frac{9}{2} \eta a \, \frac{1}{|\vec{z}|^3} \left[  \frac{\mat{I}}{3} 
     - \hat{\vec{z}}\otimes \hat{\vec{z}}  \right] \cdot \vec{v}
    \, ,
\end{multline}
moving with velocity $\vec{v}$ through a Stokes fluid with viscosity $\eta$ in $d=3$ dimensions~\cite{landau_lifshitz_1987}.
As before, $\vec{z}$ refers to the position relative to the centroid of the sphere.
In the following, I will consider a condensate consisting of many such inclusions and discuss the two opposing limits of low and high concentrations.

\subsubsection{Low concentration of inclusions}
At low concentrations, it is tempting to neglect hydrodynamic interactions among the inclusions and consider each particle as making a small additive contribution to the total pressure field.
Assuming a uniform concentration $\Delta c$ of inclusions within the condensate, the pressure gradient is given by
\begin{multline}
\label{eq:pressure_gradient_low_density}
    -\vec\nabla p = \frac{3}{2} \eta a \, \int_{\mathcal{D}} \! d^3 \vec{r} \, \Delta c \left[ \vec\nabla \otimes  \vec\nabla \left( \frac{1}{|\vec{z} - \vec{r}|} \right) \right] \cdot \vec{v} \\
    = 6\pi \eta a \, R^3 \Delta c \, \frac{1}{3}\left[ \vec\nabla \otimes  \vec\nabla \left(  \frac{1}{|\vec{z}|} \right) \right] \cdot \vec{v}
    \, .
\end{multline}
The scale of the pressure gradient will match the scale $\Delta c^2 v R^3 / M$ of the force density $-\Delta c\vec\nabla\mu$ transmitted to the fluid by the induced chemical potential [Eq.~\eqref{eq:long_range_force}, Appendix~\ref{sec:flow_interdependent}] if $M \sim \Delta c /(6\pi\eta a)$.
Making the standard estimate for the collective mobility $M \sim c /(6\pi\eta a)$ according to the Stokes-Einstein relation leads to $\Delta c \sim c$ for the characteristic concentration difference at which the capillary forces due to the chemical potential gradient will dominate.
In other words, for low concentrations of inclusions in the condensate, I expect that the long-ranged forces induced by viscous fluid stresses will dominate over translation-induced polarization.

\subsubsection{Medium to high concentration of inclusions}
At higher concentrations, hydrodynamic interactions among inclusions increase the effective viscosity of colloidal suspensions such as a biomolecular condensate~\cite{batchelor_1976}, and reduce its permeation by the solvent.
In that case, one can estimate the pressure field around the condensate by approximating it as a single impermeable sphere with radius $R$~\cite{landau_lifshitz_1987},
\begin{equation}
\label{eq:pressure_gradient_high_density}
    -\vec\nabla p 
    = -\frac{9}{2} \eta R \, \frac{1}{|\vec{z}|^3} \left[  \frac{\mat{I}}{3} 
     - \hat{\vec{z}}\otimes \hat{\vec{z}}  \right] \cdot \vec{v}
    \, .
\end{equation}
Compared to low concentrations of inclusions, the scale of the pressure gradient \emph{decreases} because in Eq.~\eqref{eq:pressure_gradient_high_density} the fluid-condensate interactions are dominated by the radius of the condensate, as opposed to the bulk volume of the condensate in Eq.~\eqref{eq:pressure_gradient_low_density}.
Now, the scale of the pressure gradient will match the scale $\Delta c^2 v R^3 / M$ of the force density $-\Delta c\vec\nabla\mu$ transmitted to the fluid by the induced chemical potential [Eq.~\eqref{eq:long_range_force}, Appendix~\ref{sec:flow_interdependent}] if $\Delta c^2 v R^3 / M \sim \frac{9}{2}\eta R v$.
Making the standard estimate for the collective mobility $M \sim c /(6\pi\eta a)$ according to the Stokes-Einstein relation leads to 
\begin{equation}
    \frac{4\pi a^3}{3} \frac{\Delta c^2}{c} \sim \frac{a^2}{R^2} \, ,
\end{equation}
for the characteristic concentration difference at which the capillary forces due to the chemical potential gradient will dominate.
Note that for $a \ll R$ this expression asymptotically decays to zero, indicating that condensate-condensate interactions will be dominated by capillary forces at medium to high concentrations.

\subsection{Screening by chemical reactions and bias of chemical reactions}
The preceding sections reported the translation-induced chemical potential profile in the limit of vanishing material turnover, $k \to 0$.
The full Poisson-Boltzmann-like theory [Eq.~\eqref{eq:continuity_nonlinear}] does not have a linear solution, although the chemical potential can be formulated via an optimization principle with an effective free energy functional. 
For small material turnover, however, the calculation so far can be easily generalized whereby $\mathcal{L}$ will be the fundamental solution of the \emph{screened} Poisson equation.
This reveals that the range of the chemical potential gradients $\vec\nabla\mu$ and all of the interactions derived from these gradients is given by the diffusive screening length, $\sqrt{M / (k\beta)}$.
A finite range of interactions will also regularize the crosstalk of condensates with distant boundaries such as the cell membrane or the nuclear membrane.

Finally, applying the translation-induced chemical potential profile [Eq.~\eqref{eq:long_range_force}, Fig.~\ref{fig::motion_induced_phenomena}A] to the Poisson-Boltzmann-like theory of condensates subject to material turnover, Eq.~\eqref{eq:continuity_nonlinear}, or the linearized Eq.~\eqref{eq:continuity}, reveals how condensate motion can affect chemical reactions.
Specifically, the chemical potential at the front of the condensate is decreased, thereby slowing down the turnover of condensate material, while the back has increased chemical potential and thus turnover of condensate material.
In conclusion, by exerting a physical force on condensates, one can reorganize chemical reactions within.

\section{Condensate motion in response to forces}
So far, the discussion centered around the prescribed motion of one phase-separated domain and how this motion can induce chemical potential gradients and hydrodynamic pressure gradients.
This raises the converse question of how externally applied fields and forces will cause condensates and phase-separated domains to move in response.
Answering this question requires a closure relation to eliminate the chemical potential from Eq.~\eqref{eq:long_range_force_comoving} and to solve for the condensate velocity $\vec{v}$.
To that end, I use the thermodynamic consistency criterion~\cite{goychuk_2024_self_consistent}, 
\begin{equation}
\label{eq:thermodynamic_consistency_criterion}
    \int_{\mathcal{D}} d^d\vec{z} \, \vec\nabla\mu(\vec{z}) = 0 \, ,
\end{equation}
where, as before, $\mathcal{D}$ is the condensate domain in the comoving frame of the condensate.
This criterion can be interpreted as the conservation of momentum for the condensate.
In other words, because the physical origin of the chemical potential profile, before dropping all the details, is a scalar free energy encoding pairwise reciprocal interactions, it cannot spontaneously break symmetry by polarizing.

After eliminating the chemical potential in Eq.~\eqref{eq:long_range_force_comoving} by integrating over the condensate and applying Eq.~\eqref{eq:thermodynamic_consistency_criterion}, one has
\begin{multline}
    0 = - \int_\mathcal{D} d^d\vec{z} \, M\vec\nabla\Psi(\vec{z}) 
    + \Delta c \, \int_\mathcal{D} d^d\vec{z} \, \mathcal{B}(\vec{z}) \, \cdot \vec{v} \\
    - \int d^d\vec{z} \, \mathcal{B}(\vec{z}) \cdot \vec{j}(\vec{z}) \, ,
\end{multline}
leaving only a substitution of the Green's tensor $\mathcal{B}(\vec{z})$ for an electrostatic dipole field, Eq.~\eqref{eq:tensor_longranged_force}.
Note that $\mathcal{B}(\vec{z})$ for $\vec{z} \in \mathcal{D}$ can also be derived for general geometries by invoking isotropy, which implies $\mathcal{B} = \tr(\mathcal{B}) \, \mat{I}/d$.

This analysis reveals without loss of generality, and independently of other microscopic details, that the centroid velocity of a condensate or any domain governed by a continuity equation [Eq.~\eqref{eq:continuity} in the limit of negligible material turnover, $k \to 0$] is given by the following \emph{force-response relation}~\cite{goychuk_2024_self_consistent, goh_2024}
\begin{multline}
\label{eq:force-response-general}
    \vec{v}(t) = \partial_t \vec{r}(t) = 
    - \frac{d}{\Delta c} \int_{\mathcal{D}(t)} \!\frac{d^d\vec{x}}{V_\mathcal{D}} \, M\vec\nabla\Psi(\vec{x},t) \\
    - \frac{d}{\Delta c} \int \!\frac{d^d\vec{x}}{V_\mathcal{D}} \, \mathcal{B}(\vec{x} - \vec{r}) \cdot  \, \vec{j}(\vec{x},t) \, ,
\end{multline}
where $\vec{z} \coloneqq \vec{x}-\vec{v}t$ and its generalization $\vec{z} \coloneqq \vec{x}-\vec{r}(t)$ were used to transform back into the laboratory frame in which the condensate moves.
Here, $V_\mathcal{D} \coloneqq \int_{\mathcal{D}} d^d\vec{z}$ is the volume and $\vec{r}(t)$ the centroid of the time-dependent domain $\mathcal{D}(t)$ of the condensate.
The last term in Eq.~\eqref{eq:force-response-general} indicates that, in contrast to gradient flow dynamics, currents that cannot be represented by the gradient of a potential generically lead to nonlocal couplings.
By combining the driven condensate motion in response to local force fields, Eq.~\eqref{eq:force-response-general}, with the understanding of how condensate motion induces dipole force fields, Eq.~\eqref{eq:long_range_force}, I will next analyze pairwise interactions between condensates in a dynamic, nonequilibrium setting.

\begin{figure}[t]
    \centering
	\includegraphics{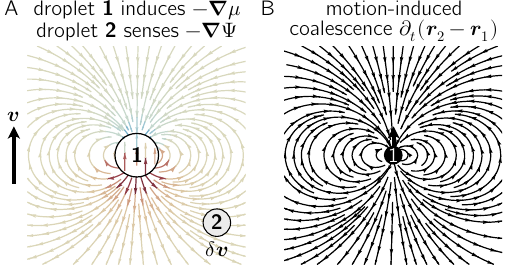}
	\caption{%
    \textbf{Interactions between two nearby condensates moving in the same direction} with a base velocity $\vec{v}$ (black arrow).
    A) One condensate (e.g., the larger one labeled ``1'') induces a chemical potential gradient $-\vec\nabla\mu_1$, while the other condensate (e.g., the smaller one labeled ``2'') senses this chemical potential gradient $-\vec\nabla\Psi_2 \equiv -\vec\nabla\mu_1$ and moves accordingly by perturbing its velocity by $\delta\vec{v}$.
    B) Condensate drift leads to directed coalescence, where smaller condensates circle around and flow (black streamlines) toward the front of larger condensates (black circle with label ``1'').
    If hydrodynamic interactions were to dominate, smaller condensates would circle toward the back of larger condensates.
	}
	\label{fig::directed_coalescence}
\end{figure}

\section{Pairs of moving condensates interact like dipoles}
The theory discussed in the preceding sections revealed that the circulatory material fluxes around moving condensates~\cite{goychuk_2024_self_consistent} are a manifestation of the dipole force field [Eq.~\eqref{eq:long_range_force}, Fig.~\ref{fig::motion_induced_phenomena}A].
The natural consequence of these fluxes, following the discussion so far, is to conclude that pairs of moving condensates must show dipole-dipole interactions according to an effective polarization along their direction of motion.
Even for passive, non-polar condensates,  these interactions emerge due to the conservation of the total mass of the phase-separating material.
Interestingly, active matter also shows dipole and quadrupole interactions, but mediated by the surrounding medium~\cite{baek_2018, ro_2021, schwarz_2002, bose_2022}, and long-range interactions can lead to clustering of self-propelling magnetic dipoles in external flow~\cite{Meng_2018}.
Motivated by these ideas, in the following I will discuss a generalized scenario in which two condensates, labeled $i\in\{1,2\}$, interact through translation-induced polarization when transported by an external field.

\subsubsection{Dynamics of the inter-condensate distance}
The main premise in this section, which is straightforward to generalize, is that each condensate moves with the same velocity $\vec{v}$ when far away from the other condensate.
This motion could be induced by phoresis, sedimentation, advection, or optical trapping, but the precise mechanism is of little consequence.
More importantly, the motion of each condensate with domain $\mathcal{D}_i(t)$ and centroid $\vec{r}_{i}(t)$ induces a perturbation in the chemical potential profile, $\delta\mu_i(\vec{x},t)$.
The resulting dipole force field perturbs the motion of the other condensate so that $\partial_t \vec{r}_{i}(t) \coloneqq \vec{v} + \delta\vec{v}_i$.
In the following, I will derive the perturbations $\delta\vec{v}_i$ [Fig.~\ref{fig::directed_coalescence}A].

According to Eq.~\eqref{eq:long_range_force}, the motion of the first condensate gives rise to an additive contribution to the chemical potential gradient,
\begin{equation}
\label{eq:chemical_potential_perturbation}
    \vec\nabla \delta\mu_1(\vec{x},t) =  
    \frac{\Delta c}{M} \, \mathcal{B}(\vec{x} - \vec{r}_1) \cdot [\vec{v} + \delta\vec{v}_1]
    \, ,
\end{equation}
where $\mathcal{B}(\vec{x} - \vec{r}_1)$ is the Green's tensor for an electrostatic dipole field [Eq.~\eqref{eq:tensor_longranged_force_general}].
The second condensate senses this perturbation of the chemical potential profile as an external potential with $\vec\nabla \Psi_2(\vec{x},t) \equiv \vec\nabla \delta\mu_1(\vec{x},t)$, which drives a potential flux.
The resulting response of the second condensate according to Eq.~\eqref{eq:force-response-general} with vanishing nonpotential flux is given by
\begin{equation}
\label{eq:velocity_perturbation}
    \delta\vec{v}_2 = - \frac{d}{\Delta c} \int_{\mathcal{D}_2} \frac{d^d\vec{z}}{V_{\mathcal{D}_2}} \, M\vec\nabla \delta\mu_1(\vec{r}_2 + \vec{z}) \, ,
\end{equation}
where $\vec{z}$ refers to a comoving coordinate system with the second condensate at its origin.

Substituting Eq.~\eqref{eq:chemical_potential_perturbation} into Eq.~\eqref{eq:velocity_perturbation} and using the definition of the coupling tensor $\mathcal{B}$ [Eq.~\eqref{eq:tensor_longranged_force_general}] leads to
\begin{equation}
    \delta\vec{v}_2 = \mat{M}_{21} \cdot [\vec{v} + \delta\vec{v}_1] \, ,
\end{equation}
with the cross-mobility simplified using Gauss' theorem,
\begin{equation}
\label{eq:mobility_matrix}
    \mat{M}_{21} \coloneqq \frac{d}{V_{\mathcal{D}_2}} \oint_{\partial\mathcal{D}_2} \!\! d\vec{S} \, \otimes \, \oint_{\partial\mathcal{D}_1} \!\! d\vec{S}' \,
    \mathcal{L}(\vec{r}_2 - \vec{r}_1 + \vec{z} - \vec{z}') \, .
\end{equation}
For small condensates that do not overlap, one can approximate $\mat{M}_{21} \approx - d \, V_{\mathcal{D}_1} \vec\nabla \otimes \vec\nabla \, \mathcal{L}(\vec{r}_2 - \vec{r}_1)$.
This result quantifies the (nonreciprocal) response of the second condensate to the motion of the first condensate; the converse response of the first condensate to the motion of the second condensate is analogous, just with flipped indices $1 \leftrightarrow 2$.

The motion of the two condensates relative to each other is quantified by $\partial_t (\vec{r}_2 - \vec{r}_1) = \delta\vec{v}_2 - \delta\vec{v}_1$.
Based on the above approximation of the cross-mobility Eq.~\eqref{eq:mobility_matrix} for small condensate sizes, one can assume that $\mat{M}_{21}$ and $\mat{M}_{12}$ commute.
The separation between the two condensates will then gradually change according to
\begin{equation}
    \delta\vec{v}_2 - \delta\vec{v}_1 \approx [\mat{I} - \mat{M}_{21} \cdot \mat{M}_{12}]^{-1} \cdot \left( 
    \mat{M}_{21} - \mat{M}_{12}
    \right) \cdot \vec{v}
    \, .
\end{equation}
For large separations between the condensates, the term in square brackets is on the order of unity.
This simplification, and substituting the cross-mobility Eq.~\eqref{eq:mobility_matrix} for small condensate sizes, leads to
\begin{equation}
    \partial_t (\vec{r}_2 - \vec{r}_1) \approx 
    d \left( 
    V_{\mathcal{D}_2}
    - V_{\mathcal{D}_1}
    \right) 
    \vec\nabla \otimes \vec\nabla \, \mathcal{L}(\vec{r}_2 - \vec{r}_1)
    \cdot \vec{v}
    \, ,
\end{equation}
for the dynamics of the condensate-condensate distance vector.

In summary, the theory predicts that condensates drifting with velocity $\vec{v}$ relative to a background concentration at rest, due to flow, sedimentation, phoresis, or other physical mechanisms, will gradually approach each other due to their emergent dipolar interactions if they have different radii $R_1 \neq R_2$.
To lowest order and contrary to an electrical dipole, a condensate will not interact with its mirror image moving parallel to an impermeable boundary.
For spherical condensates in $d=3$ dimensions, one has $\mathcal{L}(\vec{z}) = 1/(4\pi|\vec{z}|)$ and $V_{\mathcal{D}_i} = 4\pi/3 \, R_i^3$, and the relative motion of the condensate centroids $\vec{r}_i$ obeys (Fig.~\ref{fig::motion_induced_phenomena}C)
\begin{equation}
    \partial_t (\vec{r}_2 - \vec{r}_1) \approx 
    \left( R_2^3 - R_1^3 \right) 
    \vec\nabla \otimes \vec\nabla \, \frac{1}{|\vec{r}_2 - \vec{r}_1|}
    \cdot \vec{v}
    \, .
\end{equation}
In polydisperse systems, the smaller condensates will be drawn toward the front of the larger condensates.
Hence, even passive condensates undergo directed coalescence under external potential gradients.
If hydrodynamic interactions were to dominate, the flow direction would be reversed, and the smaller condensates would be drawn toward the back of the larger condensates.
The prediction of directed coalescence and front-seeking behavior can be experimentally tested on condensates in external chemical gradients~\cite{jambonpuillet_2023}, and is consistent with the observation that diffusiophoresis of biomolecules promotes phase separation~\cite{doan_2024}.

\subsubsection{Scaling analysis of directed coalescence}
How fast is directed coalescence compared to collision-coalescence by Brownian motion in $d=3$ dimensions?
To answer this question, in the following I will track $\langle R \rangle$ assuming self-similarity for the size distribution of the condensates.
Moreover, I assume that the drift velocity $\vec{v}$ is independent of the average condensate size $\langle R\rangle$, which will be justified \emph{a posteriori} by Eq.~\eqref{eq:force-response}.
With the typical distance between condensates $\langle|\vec{r}_j - \vec{r}_i|\rangle_{ij} \propto \langle{R}\rangle$, directed coalescence induces collisions with a rate $\propto\langle R\rangle^{-1}$, so that the average condensate size will increase over time as $\langle R\rangle \propto t^1$.
This result is analogous to how hydrodynamic interactions and surface tension can cooperate to increase coalescence rates during the late stages of phase separation~\cite{siggia_1979}.
If the drift velocity depends on condensate size, $\vec{v}\propto R^{-n}$, then the collision rate $\propto \langle R\rangle^{-n-1}$ and the scaling of the condensate growth is generalized to $\langle R \rangle \propto t^{1/(1+n)}$.
In contrast, for collision-coalescence by Brownian motion, the diffusion coefficient scales as $\propto R^{-1}$ according to the Stokes-Einstein relation and, over distances $\propto\langle R\rangle$, induces collisions with the rate $\propto \langle R\rangle^{-3}$, leading to the usual $\langle R\rangle \propto t^{1/3}$ scaling.
Thus, I expect that directed coalescence will dominate on long time scales.

Finally, note that the cellular nucleoplasm shows correlated flows on length scales of microns~\cite{zidovska_2013, aljord_2022}.
The theory presented so far predicts that, once condensates are within a distance of a micron of each other in the cell nucleus, directed coalescence will deterministically bring them together.
How will condensates encounter each other on larger distances?
To answer this question, the remainder of the manuscript will elucidate the Brownian motion of condensates that are far apart.

\section{Brownian motion of condensates}
The discussion so far centered on the emergence of deterministic interactions among condensates drifting under an external field relative to a background concentration at rest.
This includes, for example, advection by fluid flow, sedimentation due to gravitation, or phoresis in an electric field with parallel field lines.
On large length scales, these interactions can be neglected compared to the Brownian motion of the condensate centroid $\vec{r}(t)$.
This motion, with random velocities $\vec{v}(t) = \partial_t \vec{r}(t)$, is agitated by passive and active fluctuating forces.
Brownian motion is well characterized for colloidal particles, but not for condensates.
In the following, I will analyze the Brownian motion of condensates in a viscoelastic solvent driven by passive and active fluctuations, with a view toward describing condensates in the cell cytoplasm or nucleoplasm.

\subsection{Condensates in random fluid flows}
\subsubsection{Velocity of the condensate center}
The force-response relation, Eq.~\eqref{eq:force-response-general}, characterizes the motion of condensates due to potential and nonpotential currents.
The stochastic motion of the constituent biomolecules can be modeled, in the inviscid limit, by applying a fluctuating potential $\Psi$ that drives random thermal currents $\vec{j}_{th} \coloneqq -M\vec\nabla\Psi$ with zero mean and covariance $\langle \vec{j}_{th} (\vec{x}, t) \otimes \vec{j}_{th} (\vec{x}', t') \rangle = 2 M k_B T \mat{I} \delta(\vec{x}-\vec{x}') \, \delta(t-t')$.
This is sometimes referred to as \emph{Model B}~\cite{hohenberg_halperin_1977}.
On sufficiently large length scales, one also needs to take into account advection by stochastic hydrodynamic flows with velocity $\vec{v}_f$.
The corresponding nonpotential fluxes, $\vec{j}(\vec{x},t) \coloneqq c(\vec{x},t) \, \vec{v}_f(\vec{x},t)$, cannot be cast as a gradient.
After substituting in Eq.~\eqref{eq:force-response-general}, one has:
\begin{multline}
    \vec{v}(t) = \partial_t \vec{r}(t) = 
    \frac{d}{\Delta c} \int_{\mathcal{D}(t)} \!\frac{d^d\vec{x}}{V_\mathcal{D}} \, \vec{j}_{th}(\vec{x},t) \\
    - \frac{d}{\Delta c} \int \!\frac{d^d\vec{x}}{V_\mathcal{D}} \, c(\vec{x},t) \, \mathcal{B}(\vec{x} - \vec{r}) \cdot \vec{v}_f(\vec{x},t) \, .
\end{multline}
The integral in the second line can be split in two domains by using the sharp interface approximation for the concentration profile of the condensate, $c(\vec{x},t) = c_- + \Delta c$ for $\vec{x} \in \mathcal{D}(t)$, else $c_-$, leading to:
\begin{multline}
    \vec{v}(t) =  
    \frac{d}{\Delta c} \int_{\mathcal{D}(t)} \!\frac{d^d\vec{x}}{V_\mathcal{D}} \, \vec{j}_{th}(\vec{x},t) \\
    - d \int_{\mathcal{D}(t)} \!\frac{d^d\vec{x}}{V_\mathcal{D}} \, \mathcal{B}(\vec{x} - \vec{r}) \cdot \vec{v}_f(\vec{x},t) \\
    - \frac{d c_-}{\Delta c} \int \!\frac{d^d\vec{x}}{V_\mathcal{D}} \, \mathcal{B}(\vec{x} - \vec{r}) \cdot \vec{v}_f(\vec{x},t)
    \, .
\end{multline}
The second line can be simplified by substituting $\mathcal{B}(\vec{z}) = - \mat{I}/d$ for $|\vec{z}| \leq R$ [Eq.~\eqref{eq:tensor_longranged_force}].
Moreover, recall that the Green's tensor $\mathcal{B}$ for an electrostatic dipole field is the Hessian of the fundamental solution $\mathcal{L}$ to the Poisson equation [Eq.~\eqref{eq:tensor_longranged_force_general}], and that $\lim_{|\vec{z}|\to\infty} \vec\nabla\mathcal{L}(\vec{z}) = \vec{0}$.
Together with fluid incompressibility, $\vec\nabla \cdot \vec{v}_f = 0$, this allows eliminating the third line via integration by parts.

In summary, the centroid velocity of the condensate in incompressible fluid flow is given by
\begin{equation}
\label{eq:force-response}
    \vec{v}(t) = \frac{d}{\Delta c} \int_{\mathcal{D}(t)} \!\!\frac{d^d\vec{x}}{V_\mathcal{D}} \, \vec{j}_{th}(\vec{x}, t) 
    + \int_{\mathcal{D}(t)} \!\!\frac{d^d\vec{x}}{V_\mathcal{D}} \, \vec{v}_f(\vec{x}, t) 
    \, .
\end{equation}
The first term generalizes the result of Ref.~\cite{goh_2024} to arbitrary dimensions.
The main focus in the following is on the second term in Eq.~\eqref{eq:force-response}, which intuitively states that the condensate is advected by the local average flow of the fluid.

\subsubsection{Fluctuations of the condensate center}
The subsequent analysis will, for simplicity, neglect the feedback of condensate motion on the hydrodynamic flows.
The technical justification, presented in detail in Appendix~\ref{sec:flow_interdependent}, goes as follows.
On sufficiently large length scales and far away from impermeable physical boundaries, fluid flows induced by the long-range chemical potential gradient [Eq.~\eqref{eq:long_range_force}] will resemble those generated by a point force.
Since the gradient of the chemical potential $\vec\nabla \mu$ is of the order $\mathcal{O}(\Delta c / M)$ and the force transmitted to the fluid is $-\Delta c \vec\nabla\mu$, the resulting fluid flows are of the order $\mathcal{O}(\Delta c^2)$ and decay as $1/M$ for large mobilities $M$.
On a side note, this raises the fundamental question of how one can measure the chemical potential gradients induced by condensate motion.
Moreover, long-range chemical potential gradients induced by the random motion of condensates will directly lead to Casimir-like forces.
This is an interesting topic for future research in light of the aforementioned competition between hydrodynamics and phase separation.

For further simplicity, I will assume that the fluctuating hydrodynamic flows $\vec{v}_f$ and the thermal currents $\vec{j}_{th}$ induced by microscopic collisions are independent random variables.
The resulting Brownian motion of the condensate is characterized by the velocity [Eq.~\eqref{eq:force-response}] autocorrelation function
\begin{multline}
\label{eq:velocity-autocorrelation}
    \langle \vec{v}(t) \cdot \vec{v}(t') \rangle
    = D_c \, \delta(t-t') \\
    + \iint_{\mathcal{D}} \frac{d^d\vec{z}}{V_\mathcal{D}} \frac{d^d\vec{z}'}{V_\mathcal{D}} \, 
    \left\langle
    \vec{v}_f(\vec{r} + \vec{z}, t) \cdot \vec{v}_f(\vec{r}' + \vec{z}', t') 
    \right\rangle 
    \, ,
\end{multline}
with effective condensate diffusion coefficient $D_c \coloneqq 2d^3 M k_B T/(\Delta c^2 V_\mathcal{D})$.
Here and in the following, the shorthand notation $\vec{r} \equiv \vec{r}(t)$ and $\vec{r}' \equiv \vec{r}(t')$ will refer to the time-dependent position of the condensate center.
In the first line of Eq.~\eqref{eq:velocity-autocorrelation}, terms lacking temporal correlations were simplified using $\vec{r}(t) \sim \vec{r}(t')$ because they are nonzero only for $t=t'$.
In contrast, the second line of Eq.~\eqref{eq:velocity-autocorrelation} requires a more careful analysis because temporal correlations can effectively quench the flow landscape navigated by the condensate.

\subsubsection{Condensates measure random flows at random locations}
The difficulty in analyzing Eq.~\eqref{eq:velocity-autocorrelation} is that not only the fluid flow but also the reference frame of the condensate with center $\vec{r}(t)$ which senses this flow are time-dependent random variables.
To better understand the consequences of this geometric coupling, consider a Taylor series expansion of the fluid flow velocity field around the midpoint $\bar{\vec{r}} \coloneqq (\vec{r} + \vec{r}')/2$ of the condensate trajectory in the time window $[t,t']$,
\begin{equation}
\label{eq:flow_expansion_instantaneous}
    \vec{v}_f(\vec{z}+\vec{r},t) 
    = \sum_n \frac{1}{n!} [(\vec{r} - \bar{\vec{r}}) \cdot \vec\nabla]^n \vec{v}_f(\vec{z} + \bar{\vec{r}}, t) \, ,
\end{equation}
where $\vec{r} \equiv \vec{r}(t)$ and $\vec{r}' \equiv \vec{r}(t')$ are random variables corresponding to the centroid of the condensate.

This expansion reveals a multiplicative coupling between two random variables, namely the centroid of the condensate and the fluid flow, which requires careful deliberation to treat analytically.
According to Eq.~\eqref{eq:force-response} the centroid motion of the condensate is determined by a thermal flux (first term), which is taken statistically independent of the fluid flow ${\langle\vec{j}_\textit{th}\cdot\vec{v}_f\rangle = 0}$, plus the average fluid flow in the condensate (second term).
The spatial averaging cancels out the random fluid flows in the condensate, rendering the mean flow much slower than the flow in a small fluid parcel.
This allows approximating ${V_{\mathcal{D}}^{-1}\int_{\mathcal{D}}\!d^d\vec{z} \, \langle \vec{v}_f(\vec{z},t) \cdot \vec{v}_f(\vec{z}',t')\rangle \approx 0}$ unless fluid flows are correlated over large distances.
Taken together, the condensate position $\vec{r}(t) = \int_{-\infty}^t \! d\tau \, \vec{v}(\tau)$ fluctuates much less than the fluid flow, $\langle \vec{r}(t) \cdot \vec{v}_f(\vec{z},t') \rangle \approx 0$.
This enables a preaveraging approximation, $\vec{v}_f(\vec{z} + \bar{\vec{r}},t) \cdot \vec{v}_f(\vec{z}' + \bar{\vec{r}},t') \approx \langle \vec{v}_f(\vec{z} + \bar{\vec{r}},t) \cdot \vec{v}_f(\vec{z}' + \bar{\vec{r}},t') \rangle$.
The above argument breaks down if the fluid flows are correlated over large distances.
In that case, however, the dominant contribution to Eq.~\eqref{eq:flow_expansion_instantaneous} is the $n=0$ term so that position-flow correlations are irrelevant and the preaveraging approximation does not affect the results.

Together with these approximations, the series expansion allows working out the correlations between the fluid flow in a fluctuating frame of reference, as detailed in Appendix~\ref{sec:flow_expansion} by applying Wick's theorem,
\begin{multline}
\label{eq:flow_expansion}
    \langle\vec{v}_f(\vec{r} + \vec{z},t) \cdot \vec{v}_f(\vec{r}' + \vec{z}',t')\rangle \\
    \approx \exp\biggl[
    \frac{1}{2d}
    \left\langle 
    [\vec{r}  - \vec{r}']^2
    \right\rangle  
    \vec\nabla_{\vec{z}}^2
    \biggr] \, \mathcal{C}_f(\vec{z} - \vec{z}', t - t')
    \, ,
\end{multline}
where $\mathcal{C}_f(\vec{z} - \vec{z}', t - t') \coloneqq\langle\vec{v}_f(\vec{z}, t) \cdot \vec{v}_f(\vec{z}', t') \rangle$ quantifies the hydrodynamic fluctuations in the laboratory frame.
In essence, this result describes how the stochastic motion of the condensate effectively smears out the observed correlations in the fluid flow.
This approximation will be used in the following to study how fluid flows with finite correlation times drive the Brownian motion of condensates.

\subsubsection{Mean-squared displacement}
It is more convenient to track the mean squared displacement of the condensate, $\MSD(t-t') \coloneqq \langle |\vec{r}(t) - \vec{r}(t')|^2\rangle$, instead of its velocity autocorrelation function, $\frac{1}{2}\partial_t^2 \MSD(t-t') = \langle \vec{v}(t) \cdot \vec{v}(t') \rangle$.
The memory effects arising from the interplay between condensate motion and hydrodynamic fluctuations, as seen by substituting Eq.~\eqref{eq:flow_expansion} in Eq.~\eqref{eq:velocity-autocorrelation}, lead to the following second-order differential equation:
\begin{multline}
\label{eq:ode_transformed_final}
    \frac{1}{2} \partial_t^2  
    \MSD(t)  
    = D_c \delta(t) \\
    + \left\langle
    \exp\biggl[
    \frac{\MSD(t)}{2d}   
    \vec\nabla^2
    \biggr] \mathcal{C}_f(\vec{z} - \vec{z}', t) \right\rangle_{\mathcal{D}}
    ,
\end{multline}
%
%
%\begin{multline}
%\label{eq:ode_transformed_final}
%    \frac{1}{2} \partial_t^2  
%    \MSD(t)  
%    = \\
%    = \iint_{\mathcal{D}} \frac{d^d\vec{z}}{V_\mathcal{D}} \frac{d^d\vec{z}'}{V_\mathcal{D}}   
%    %
%    \exp\biggl[
%    \frac{\MSD(t)}{2d}   
%    \vec\nabla^2
%    \biggr] \mathcal{C}_f(\vec{z} - \vec{z}', t) 
%    %
%    ,
%\end{multline}
%
%defined for , 
where $\langle \dots \rangle_{\mathcal{D}} \coloneqq V_{\mathcal{D}}^{-2}\iint_{\mathcal{D}}\! d^d \vec{z} \, d^d \vec{z}' \dots$ indicates spatial averaging over the coordinates $\vec{z}$ and $\vec{z}'$ inside the phase-separated domain $\mathcal{D}$.
This can be solved numerically, for spherical condensates with radius $R$, by using the Fourier transform $\widehat{\mathcal{C}}_f(|\vec{q}|, \omega) \coloneqq \int d^d\vec{z} \int dt \, e^{-i\vec{q}\cdot\vec{z} - i\omega t} \, \mathcal{C}_f(|\vec{z}|, t)$ of the hydrodynamic fluctuations.
The Fourier-transformed dynamics is presented in Appendix~\ref{sec:velocity_autocorrelation_equation}.
Restricting the time domain to positive times $t > 0$ allows integrating out the $\delta$-distributions in the dynamics and leads to the two initial conditions $\MSD(0) = 0$ and $\partial_t \MSD(0^{+}) = D_0$, where
\begin{equation}
\label{eq:ode_transformed_boundary_condition}
    D_0 = D_c 
    + \frac{d \,\Gamma\bigl(1+\tfrac{d}{2}\bigr)}{\pi^{d/2} R^d} 
    \int_0^{\infty} \! \frac{dq}{q} \,   
    J_{d/2}^2(q R) \, 
    \widehat{\mathcal{C}}_f(q, \infty) \, ,
\end{equation}
characterizes the fluctuations of the condensate center of mass on short time scales due to temporally independent kicks.
Here, the Bessel function of the first kind, $J_{d/2}$, accounts for the spherical geometry of the condensate.

\begin{figure}[h]
    \centering
	\includegraphics{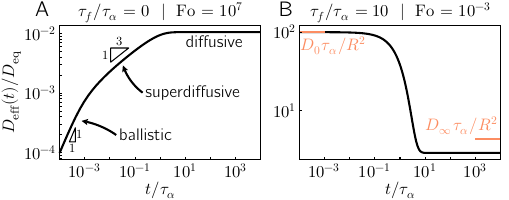}
	\caption{%
    Effective diffusion coefficient $D_\text{eff}(t) \coloneqq \partial_t \MSD(t)$ of a condensate in a viscoelastic medium compared to Brownian motion $D_\text{eq} \coloneqq 6 k_B T/(5\pi \eta R)$ in an equilibrium-like setting. 
    Numerical results (black solid lines) were obtained by solving Eq.~\eqref{eq:ode_transformed_final} with fluid flow correlations given by Eq.~\eqref{eq:fluid_correlation} for negligible hydrodynamic screening $\gamma \approx 0$.
    A) For $\tau_\alpha \gg \tau_f$ and high Fourier numbers $\text{Fo}$ [Eq.~\eqref{eq:parameter_estimation_fourier_number}], one recovers an intermediate scaling regime indicating superdiffusion with $\MSD(t) \sim t^{4/3}$ as predicted for viscous fluids in a quenched flow landscape.
    On short time scales, one finds ballistic motion, whereas on long times one recovers regular diffusion.
    B) For $\tau_\alpha < \tau_f$ and small Fourier numbers $\text{Fo}$, diffusion on long time scales is much slower than fluctuations on short time scales.
    Thus, for this parameter regime, there is no superdiffusive transient.
    Red lines show asymptotic approximations for the diffusion coefficients according to Eq.~\eqref{eq:ode_transformed_boundary_condition} and Eq.~\eqref{eq:diffusion_coefficient_long}.
	}
	\label{fig::effective_diffusion_coefficient_appendix}
\end{figure}

\subsubsection{Anomalous diffusion}
To better understand the memory encoded in Eq.~\eqref{eq:ode_transformed_final}, I will briefly discuss the dynamics on time scales shorter than the correlation time of the fluid.
To that end, consider advection in a quenched landscape characterized by $\widehat{\mathcal{C}}_f(q, \omega) = q^n \delta(\omega)$.
In this case, Eq.~\eqref{eq:ode_transformed_final} is equivalent to $\partial_t^2 \MSD \sim \MSD^{-(d+n)/2}$ for $n>-d$ which leads to $\MSD \sim t^{4/(2+d+n)}$ after a scaling ansatz.
In $d=3$ dimensions, this corresponds to superdiffusion with exponent $4/3$ for quenched hydrodynamic flows ($n=-2$, see also Fig.~\ref{fig::effective_diffusion_coefficient_appendix}A based on the viscoelastic flows discussed later), or subdiffusion with exponent $4/5$ for motion in a quenched random potential landscape ($n=0$).
Hence, phase-separated domains can show anomalous diffusion on short to intermediate time scales; please refer to Ref.~\cite{bouchaud_1990} for an in-depth discussion of anomalous diffusion.

\subsubsection{Dynamics on long time scales}
To characterize the Brownian motion of the condensate on long time scales, it is useful to define the effective diffusion coefficient $D_\infty \coloneqq \lim_{t\to\infty} \partial_t \MSD(t)$.
The key idea is that the exponential kernel in Eq.~\eqref{eq:ode_transformed_final} decays to zero on sufficiently long time scales, which allows integrating it out,
\begin{equation}
%\label{eq:diffusion_coefficient_long}
    D_\infty  
    = D_c + 
    \int_{-\infty}^{\infty} dt\left\langle
    \exp\biggl[
    \frac{\MSD(t)}{2d}   
    \vec\nabla^2
    \biggr] \mathcal{C}_f(\vec{z} - \vec{z}', t) \right\rangle_{\mathcal{D}}
    .
\end{equation}
%
%
%\begin{multline}
%\label{eq:diffusion_coefficient_long}
%    D_\infty  
%    \approx D_0
%    + \frac{d \,\Gamma\bigl(1+\tfrac{d}{2}\bigr)}{\pi^{d/2} R^d}
%    \int_0^{\infty} \! \frac{dq}{q} \,
%    J_{d/2}^2(q R) \,  
%    \color{lightgray}\times\color{black} \\ \color{lightgray}\times\color{black}  
%    \int \frac{d\omega}{2\pi}
%    \frac{4 d D_0 q^2}{4 d^2 \omega ^2 + D_0^2 q^4}  
%    \left[\widehat{\mathcal{C}}_f(q, \omega) - \widehat{\mathcal{C}}_f(q, \infty)\right]
%    %
%    \, ,
%\end{multline}
%
In the following, I make the approximation that the smearing out of the fluid flow sensed by the condensate is dominated by condensate motion on very short time scales, $\MSD(t) \approx D_0 \, |t|$.
This is an excellent approximation if the effective diffusion coefficient of the condensate, $D_\text{eff}(t) \coloneqq \partial_t \MSD(t)$, is large so that the squared distance $\MSD(t)$ traveled within the decorrelation time of the fluid is large compared to the typical length scale of the flow correlations.
In summary, one has
\begin{equation}
\label{eq:diffusion_coefficient_long}
    D_\infty  
    \approx D_c + 
    \int_{-\infty}^{\infty} dt\left\langle
    \exp\biggl[
    \frac{D_0 |t|}{2d}   
    \vec\nabla^2
    \biggr] \mathcal{C}_f(\vec{z} - \vec{z}', t) \right\rangle_{\mathcal{D}}
    ,
\end{equation}
which allows making analytical estimates (see Appendix~\ref{sec:asymptotics} for the corresponding Fourier space representation).
The dynamics on long time scales can be either faster, or actually slower than the motion on very short time scales determined by $D_0$.

The intuition for the theoretical framework discussed so far is as follows.
Without temporal correlations in the fluid flow, the motion of the condensate from coordinate $\vec{r}(t)$ to $\vec{r}(t')$ is accompanied by a complete loss of information about the flow landscape.
In that case, the dynamics of the condensate motion will resemble an equilibrium system (\emph{equilibrium-like}), with diffusion coefficient $D_0 = D_\infty$ given by Eq.~\eqref{eq:ode_transformed_boundary_condition}.
In contrast, in the case of a persistent fluid flow landscape, the condensate has a memory of its past trajectory.
This memory can either amplify the motion of the condensate by ``surfing'' along the most likely currents or suppress transport by ``trapping'' the condensate.
In summary, the theory describes how the memory introduced by active forces, for example, due to cytoskeletal stirring, will affect the dynamics of condensates.

\begin{figure}[t]
    \centering
	\includegraphics{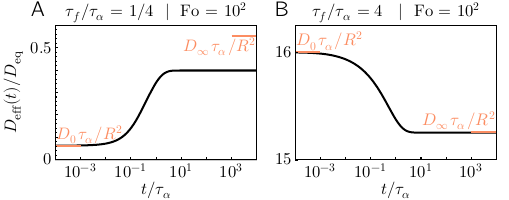}
	\caption{%
    Effective diffusion coefficient $D_\text{eff}(t) \coloneqq \partial_t \MSD(t)$ of a condensate in a viscoelastic medium compared to Brownian motion $D_\text{eq} \coloneqq 6 k_B T/(5\pi \eta R)$ in an equilibrium-like setting.
    Numerical results (black solid lines) were obtained by solving Eq.~\eqref{eq:ode_transformed_final} with fluid flow correlations given by Eq.~\eqref{eq:fluid_correlation} for negligible hydrodynamic screening $\gamma \approx 0$.
    A) For $\tau_\alpha > \tau_f > 0$, condensate motion is suppressed.
    Smaller values of $\tau_f / \tau_\alpha$ further reduce the growth rate of $\MSD(t)$ and hence the decay of $\exp[-q^2 \MSD(t) / 6]$ in Eq.~\eqref{eq:ode_transformed_final}.
    This affects the fidelity of the asymptotic limits [Eq.~\eqref{eq:ode_transformed_boundary_condition} and Eq.~\eqref{eq:diffusion_coefficient_long}, red lines] compared to the numerical solution (black solid lines).
    B) For $\tau_\alpha < \tau_f$, condensate motion is enhanced relative to an equilibrium-like system satisfying the fluctuation-dissipation relation.
    For the parameters chosen here, the diffusivity depends only weakly on the time scale.
    In both panels, the Fourier number, which relates the stirring time to the characteristic time of diffusive mass transport, is $\text{Fo} = D_\text{eq} \tau_\alpha/R^2 = 100$.
	}
	\label{fig::effective_diffusion_coefficient}
\end{figure}

\subsection{Condensates in an active Maxwell fluid}
The general theory discussed so far is agnostic to the material properties of the condensate and the solvent.
Inspired by the mechanobiology of the cell, I will now apply this framework to study condensates in the fluctuating nucleoplasm or cytoplasm, which couple phase separation with fluctuating hydrodynamics~\cite{landau_lifshitz_1987, cates_review_2018, barker_2023, zhang_2024_brownian, bell_2025_comment}.
To this end, the following analysis will start by discussing the hydrodynamic fluctuations, encoded in the correlation function $\mathcal{C}_f$, for an incompressible viscoelastic fluid with hydrodynamic screening borrowed from the Brinkman model~\cite{brinkman_1949}.
For unsteady flow at low Reynolds numbers, the corresponding local balance of forces reads
\begin{equation}
\label{eq:fluid_force_balance}
    \rho \, \partial_t \vec{v}_f = \vec\nabla \cdot \mat\sigma - \vec\nabla p - \eta\gamma \vec{v}_f + [\vec\nabla \cdot \mat\alpha - c \vec\nabla\mu] \, ,
\end{equation}
where $\rho$ is the density of the fluid.
The pressure $p$ enforces incompressibility, $\vec\nabla \cdot \vec{v}_f = 0$, similar to how the chemical potential $\mu$ enforces droplet cohesion in Eq.~\eqref{eq:continuity}. 
The hydrodynamic screening by other inclusions such as large organelles is quantified by the parameter $\gamma$.

Viscoelastic effects are incorporated through the Maxwell model,
$\tau_f \partial_t\mat\sigma + \mat\sigma = \eta \, [\vec\nabla\otimes\vec{v}_f + (\vec\nabla\otimes\vec{v}_f)^T ]$,
with viscosity $\eta$ and elastic relaxation time $\tau_f$, which I envision as an idealization of chromatin.
As an alternative to this simple model of the nucleoplasm (or cytoplasm), one could also use a more detailed two-fluid description of chromatin hydrodynamics~\cite{bruinsma_2014} or directly use flow correlation functions extracted from experimental data~\cite{zidovska_2013}.
Analogously, the same framework can be applied in separate contexts to study other complex fluids or even solids such as the Kelvin-Voigt model.

The term in square brackets in Eq.~\eqref{eq:fluid_force_balance} can be identified as an applied force field, which maps to the resulting fluid flows via the following Fourier-transformed Green's function,
\begin{equation}
\label{eq:greens_function}
    \widehat{\mathcal{G}}(\vec{q}, \omega) = \frac{1}{\eta}
    \left[\frac{q^2}{1 + i\tau_f\omega} + \gamma + i \omega \frac{\rho}{\eta} \right]^{-1}
    \left[ \mat{I}- \frac{\vec{q}\otimes \vec{q}}{q^2} \right] .
\end{equation}
In the following, for simplicity, the discussion is based on an approximation for quasi-steady flows, $\omega\rho/\eta \approx 0$, which precludes studying the effect of Basset forces.

The fluid flows are driven by the random stress tensor $\mat\alpha$ which can have passive and active parts.
To interpolate between equilibrium-like and active stresses in the steady-flow approximation $\omega\rho/\eta \approx 0$, I will consider a random stress tensor with exponential memory, 
\begin{multline}
\label{eq:stress_covariance}
    \langle\alpha_{ij}(\vec{x},t) \alpha_{kl}(\vec{x}',t')\rangle = \frac{k_B T \eta}{\tau_\alpha} \, e^{-\frac{|t-t'|}{\tau_\alpha}} \, \delta(\vec{x} - \vec{x}')  \color{gray}\times\color{black} \\ \color{gray}\times\color{black}
    \left[\delta_{ik}\delta_{jl} + \delta_{il}\delta_{jk} - \frac{2}{d} \delta_{ij}\delta_{kl}\right] \, .
\end{multline}
The main features of these stresses are a 4-tensor correlation structure that eliminates the isotropic part (fluctuating pressure) while keeping the deviatoric shear stress along independent directions, a lack of spatial correlations, and temporal correlations on a time scale $\tau_\alpha$.
In the limit of vanishing stress correlation time (also referred to as stirring time), $\tau_\alpha \to 0$, this converges to the fluctuating hydrodynamic stresses expected in a Stokes fluid~\cite{zwanzig_1964_fluctuating_hydro}, which has vanishing relaxation time $\tau_f \to 0$.
From the perspective of the generalized Langevin equation, the scenario $\tau_\alpha = \tau_f$ satisfies the fluctuation-dissipation theorem~\cite{Kubo_1966} and therefore corresponds to an equilibrium-like system.
The ratio $\tau_\alpha / \tau_f$ between these time scales provides a handle on tuning the nonequilibrium dynamics of the viscoelastic fluid and the embedded condensate.

As discussed previously, condensates do not significantly perturb the fluid flow if the mobility $M$ is large or the concentration jump $\Delta c$ at the droplet boundary is small.
Relaxing this approximation (Appendix~\ref{sec:flow_interdependent}) would enable the calculation of fluctuation-induced interactions between condensates, which is reserved for future work.
The linear response [Eq.~\eqref{eq:greens_function}] to the active stress fluctuations [Eq.~\eqref{eq:stress_covariance}] predicts that the velocity fluctuations in the viscoelastic fluid are characterized by
\begin{equation}
\label{eq:fluid_correlation}
\frac{\eta \, \widehat{\mathcal{C}}_f(\vec{q}, \omega)}{k_B T} =  
\frac{2 (d - 1) \, q^2}{(q^2 + \gamma)^2 + \gamma^2 \tau_f^2 \omega^2} 
\frac{1+\tau_f^2 \omega ^2}{1+\tau_\alpha^2\omega^2}
\, .
\end{equation}
Note that for spatially uncorrelated \emph{active forces} with exponential memory, as opposed to \emph{active stresses} [Eq.~\eqref{eq:stress_covariance}], this expression would be modified by dividing by $q^2$.
In this case, the growth rate of the mean squared displacement according to Eq.~\eqref{eq:ode_transformed_final} would diverge as $\propto \gamma^{-2 + d/2}$ on short time scales.
Physically, this implies that in $d < 4$ dimensions and in the absence of hydrodynamic screening, $\gamma \to 0$, spherical condensates are always unstable with respect to active forces, leading to condensate splitting or shape fluctuations.

\begin{figure}[t]
    \centering
	\includegraphics{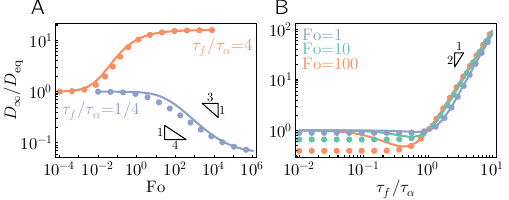}
	\caption{%
    Diffusion coefficient of condensates on long time scales, relative to an equilibrium-like setting ($\tau_f = \tau_\alpha$), as a function of A) the Fourier number $\text{Fo}$ [Eq.~\eqref{eq:parameter_estimation_fourier_number}] and B) the ratio between the fluid relaxation time $\tau_f$ and the stress correlation time $\tau_\alpha$.
    Solid lines represent analytical approximations [Eq.~\eqref{eq:diffusion_coefficient_long}], while symbols represent numerical calculations.
    Numerical results were obtained by solving Eq.~\eqref{eq:ode_transformed_final} with fluid flow correlations given by Eq.~\eqref{eq:fluid_correlation} for negligible hydrodynamic screening $\gamma \approx 0$.
	}
	\label{fig::diffusion_coefficient_comparison}
\end{figure}

\subsection{Active stirring can enhance or suppress motion}
To make concrete predictions, I will now focus the theory on spherical condensates with radius $R$ in $d=3$ dimensions.
Moreover, consider a simplified scenario in which the screening length is large compared to the condensate size, $\sqrt{\gamma} R \ll 1$, on sufficiently long time scales so that $\gamma R^2 \tau_f /t \ll 1$.
With these approximations, the flow correlations of the viscoelastic fluid [Eq.~\eqref{eq:fluid_correlation}] can be simplified to
\begin{equation}
\label{eq:fluid_correlation_approximated}
\frac{\eta \, \widehat{\mathcal{C}}_f(\vec{q}, \omega)}{k_B T} \approx  
\frac{4 \, q^2}{q^4 + 2 \gamma q^2} 
\frac{1+\tau_f^2 \omega ^2}{1+\tau_\alpha^2\omega^2}
\, .
\end{equation}
For $\tau_\alpha = \tau_f$, which satisfies the fluctuation-dissipation theorem, the last term in Eq.~\eqref{eq:fluid_correlation_approximated} vanishes.
Hence, the Brownian motion of condensates on long time scales can only differ from the fluctuations on short time scales, $D_0 \neq D_\infty$, if $\tau_\alpha \neq \tau_f$.

In the following, I used Eq.~\eqref{eq:fluid_correlation_approximated} to make analytical predictions based on the theoretical framework discussed so far.
Moreover, I tested these predictions by numerically integrating Eq.~\eqref{eq:ode_transformed_final}, with fluid flow correlations given by Eq.~\eqref{eq:fluid_correlation} with $\gamma = 0$, to obtain the mean squared displacement of the condensate.

\subsubsection{Fluctuations on short time scales}
Figure~\ref{fig::effective_diffusion_coefficient} shows the slope $D_\text{eff}(t) \coloneqq \partial_t \MSD(t)$ of the numerically calculated mean squared displacement for two scenarios, $\tau_\alpha < \tau_f$ and $\tau_\alpha > \tau_f$.
On short time scales, the mean squared displacement of the condensate grows with the effective diffusion coefficient
\begin{equation}
\label{eq:dynamics_3d_short_time scale}
    D_0 \approx D_c 
    + \frac{6 k_B T}{5\pi \eta R} 
    \frac{\tau_f^2}{\tau_\alpha^2}
    \left[1-\frac{5 \sqrt{\gamma} R}{3\sqrt{2}}\right] 
    \, ,
\end{equation}
to lowest order in $\sqrt{\gamma}R$ , which was derived from Eq.~\eqref{eq:ode_transformed_boundary_condition} with fluid flow correlations given by Eq.~\eqref{eq:fluid_correlation_approximated}.
For condensates smaller than some threshold size, $R \ll R^\star$, motion is dominated by $D_c \propto 1/R^3$, and is independent of fluctuations in the viscoelastic fluid.
As will be discussed in the next section, one can assume that biomolecular condensates exceed this threshold size $R^\star$ in practice.
Therefore, the subsequent discussion will be based on the converse approximation $D_c \approx 0$.
For large condensates, $R \gg R^\star$, taking into account fluid stresses recovers the size dependence of the Stokes-Einstein relation, $D_0 \propto 1/R$, in agreement with recent Cahn-Hilliard-Navier-Stokes simulations~\cite{zhang_2024_brownian}.
Compared to a solid object, the hydrodynamic radius of liquid biomolecular condensates is reduced by a factor of $5/36$ due to permeation by the fluid.
Importantly, shorter stress correlation times $\tau_\alpha$, or longer fluid relaxation times $\tau_f$, amplify the center-of-mass fluctuations of condensates relative to an equilibrium-like system ($\tau_\alpha = \tau_f$).

\subsubsection{Parameter estimation}
\label{sec:parameter_estimate}

In Eq.~\eqref{eq:dynamics_3d_short_time scale} describing condensate fluctuations on short timescales, the dimensionless ratio
\begin{equation}
\label{eq:parameter_estimation_dimensionless_ratio}
    \frac{6 k_B T}{5\pi \eta R \, D_c} = \frac{2 \Delta\phi^2 R^2}{15 a^2} \, ,
\end{equation}
determines which term is dominant. 
To estimate $D_c = 2d^3 M k_B T/(\Delta c^2 V_\mathcal{D})$, I have identified $\nu^{-1}$ as a typical concentration scale, with $\nu = \frac{4\pi}{3} a^3$ the molecular volume of a particle with radius $a\sim\SI{2.5}{\nano\meter}$. 
According to the Stokes-Einstein relation, the corresponding collective mobility then is $M = \nu^{-1}/(6\pi\eta a)$.
Moreover, I have defined the volume fraction difference $\Delta\phi \coloneqq \Delta c \, \nu$ across the interface of the condensate.
Assuming a volume fraction difference of $\Delta\phi = 0.2$ for phase-separating biomolecules, the second term in Eq.~\eqref{eq:dynamics_3d_short_time scale} will dominate if the condensate is larger than
\begin{equation}
    R \gg R^\star = \sqrt{\frac{15}{2}} \frac{a}{\Delta\phi} \sim \SI{35}{\nano\meter} \, .
\end{equation}
Note that if one chooses a lower characteristic concentration for estimating the collective mobility, such as $0.2 \nu^{-1}$ for a volume fraction of $0.2$, then $M$ and $D_c$ will be lower, increasing the left-hand side of Eq.~\eqref{eq:parameter_estimation_dimensionless_ratio} and reducing $R^\star$.
In summary, the critical condensate size $R^\star$ is only a little bit larger than the size of the phase-separating molecules, and condensates can be assumed to exceed this size for all practical purposes.

It is also useful to define a dimensionless quantity that relates the stirring time $\tau_\alpha$ to the characteristic time of mass transport by diffusion $D_0$ over the length scale $R$ of the condensate, in a setting resembling thermal equilibrium ($\tau_\alpha = \tau_f$).
This will be referred to as the Fourier number, 
\begin{equation}
\label{eq:parameter_estimation_fourier_number}
    \text{Fo} \coloneqq \frac{6 k_B T \tau_\alpha}{5\pi \eta R^3} \, .
\end{equation}
For a condensate with radius $R = \SI{100}{\nano\meter}$, and a stirring time of $\tau_\alpha = \SI{1}{\second}$~\cite{zidovska_2013}, the Fourier number is $\text{Fo} \sim 1600$.
For a condensate with radius $R = \SI{1}{\micro\meter}$, the Fourier number is $\text{Fo} \sim 1.6$.
Taken together, this defines the rough parameter space applicable to biomolecular condensates.

\begin{figure}[t]
    \centering
	\includegraphics{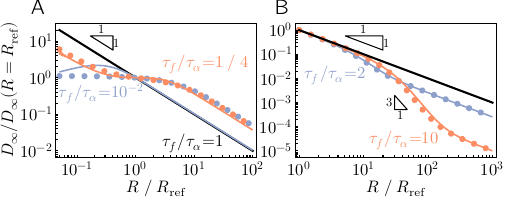}
	\caption{%
    Effective diffusion coefficient of a condensate with radius $R$, compared to a reference condensate with radius $R_\text{ref}$.
    Solid lines indicate analytical approximations [Eq.~\eqref{eq:diffusion_coefficient_long}], while symbols represent numerical calculations, where $\text{Fo} = 1000$ for the reference radius $R_\text{ref}$.
    Numerical results were obtained by solving Eq.~\eqref{eq:ode_transformed_final} with fluid flow correlations given by Eq.~\eqref{eq:fluid_correlation} for negligible hydrodynamic screening $\gamma \approx 0$.
    A) For $\tau_\alpha > \tau_f$, there is a crossover regime where the diffusion coefficient is largely independent of condensate radius.
    Note that the analytical approximation fails when $\tau_f \ll \tau_\alpha$.
    B) For $\tau_\alpha < \tau_f$, there is a crossover regime where the diffusion coefficient departs from the Stokes-Einstein scaling.
    The width of this crossover regime increases with $\tau_f / \tau_\alpha$.
	}
	\label{fig::diffusion_coefficient_size}
\end{figure}

\subsubsection{Brownian motion on long time scales}
Next, I will discuss the Brownian motion of the condensate on long time scales [Eq.~\eqref{eq:diffusion_coefficient_long}], to lowest order in $\sqrt{\gamma}R$.
For large condensates, $R \gg \sqrt{D_0 \tau_\alpha}$, one approximately has [Eq.~\eqref{eq:diffusion_coefficient_long} with fluid flow correlations given by Eq.~\eqref{eq:fluid_correlation_approximated}]
\begin{equation}
\label{eq:dynamics_3d_long_time scale}
    D_\infty  
    \approx D_0
    + \frac{6k_B T}{5\pi \eta R} \left[1 - \frac{\tau_f^2}{\tau_\alpha^2}\right]
    \left[1 - \frac{5 \sqrt{\gamma} R}{3 \sqrt{2}} \right]
    \, .
\end{equation}
Substituting Eq.~\eqref{eq:dynamics_3d_short_time scale} into Eq.~\eqref{eq:dynamics_3d_long_time scale} eliminates the dependence on $\tau_\alpha / \tau_f$, thus suggesting that large condensates show equilibrium-like dynamics on long time scales.
Together with the previous results, this indicates that condensate motion on long time scales ($D_\infty$) is faster than condensate motion on short time scales ($D_0$) for longer stress correlation times $\tau_\alpha$ than the fluid relaxation times $\tau_f$ (Fig.~\ref{fig::effective_diffusion_coefficient}A).
For $\tau_\alpha \gg \tau_f$, one has $D_0 \to 0$ and the dynamics on short time scales becomes ballistic (Fig.~\ref{fig::effective_diffusion_coefficient_appendix}A).
Conversely, condensate motion on long time scales ($D_\infty$) is slower than on short time scales ($D_0$) for $\tau_\alpha < \tau_f$ (Fig.~\ref{fig::effective_diffusion_coefficient}B and Fig.~\ref{fig::effective_diffusion_coefficient_appendix}B).
These analytical predictions are also corroborated by Fig.~\ref{fig::diffusion_coefficient_comparison}, which directly compares the numerical results with analytical approximations.

Interestingly, the scaling of the effective diffusion coefficient with condensate size changes qualitatively for sufficiently small condensates, $R \ll \sqrt{D_0 \tau_\alpha}$.
The analytical approximation,
\begin{equation}
\label{eq:dynamics_3d_long_time scale_small_condensate}
    D_\infty  
    \approx D_0
    + \frac{\sqrt{6} k_B T}{\pi \eta \sqrt{D_0 \tau_\alpha}} \left[1 - \frac{\tau_f^2}{\tau_\alpha^2}\right]
    \left[1-\sqrt{\frac{\gamma D_0 \tau_\alpha}{3}}\right]
    \, ,
\end{equation}
suggests a potential nonmonotonicity as a function of $D_0$ and thus condensate radius if $\tau_\alpha > \tau_f$.
The numerical calculations clarify that, instead of nonmonotonicity, there is a wide range of condensate sizes where the diffusion coefficient on long time scales, $D_\infty$, is actually insensitive to the condensate radius (Fig.~\ref{fig::diffusion_coefficient_size}A).
Hence, active stirring of the viscoelastic medium qualitatively changes the Brownian motion of condensates, with potentially far-reaching implications for coalescence.
These findings complement the prediction of directed coalescence as a result of translation-induced dipole interactions, which was discussed in the first part of this manuscript.

\subsubsection{Scaling analysis of Brownian collision-coalescence}
\begin{figure}[t]
    \centering
	\includegraphics{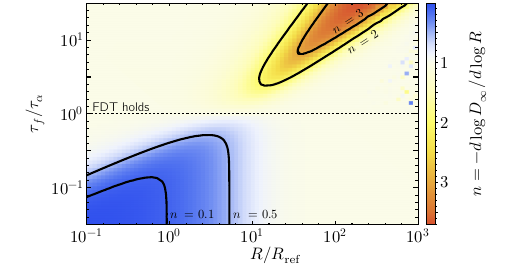}
	\caption{%
    Scaling of the effective diffusion coefficient with condensate radius, $D_\infty \sim 1/R^n$, compared to a reference condensate with radius $R_\text{ref}$ and corresponding Fourier number $\text{Fo} = 1000$.
    In the blue region of the phase diagram, condensate motion is almost independent of condensate size.
    In the red region of the phase diagram, the dependence of Brownian motion on condensate size is amplified.
    In all other regions (bright yellow), the Stokes-Einstein relation $D_\infty \sim 1/R$ is recovered.
    Dashed line corresponds to a passive system in which the fluctuation-dissipation theorem (FDT) holds.
    Numerical results were obtained by solving Eq.~\eqref{eq:ode_transformed_final} with fluid flow correlations given by Eq.~\eqref{eq:fluid_correlation} for negligible hydrodynamic screening $\gamma \approx 0$.
	}
	\label{fig::diffusion_size_scaling}
\end{figure}

How will active stirring affect the growth of the condensate radii via collision-coalescence?
For $\tau_\alpha > \tau_f$, active fluctuations turn the effective diffusion coefficient $D_\infty$ independent of condensate size (Fig.~\ref{fig::diffusion_coefficient_size}A, Fig.~\ref{fig::diffusion_size_scaling}).
This increases the scaling exponent of the average condensate radius as a function of time, $\langle R \rangle \sim t^{1/2}$.
Once the condensates are sufficiently large, the effective diffusion coefficient recovers the Stokes-Einstein size dependence $D_\infty \propto 1/R$ so that further growth follows the slower, standard $\langle R \rangle \sim t^{1/3}$ law.
For $\tau_\alpha < \tau_f$, the effective diffusion coefficient scales as $D_\infty \propto 1/R^n$, with $n>1$, over a wide range of condensate sizes (Fig.~\ref{fig::diffusion_coefficient_size}B, Fig.~\ref{fig::diffusion_size_scaling}), leading to a reduced scaling exponent with $\langle R \rangle \sim t^{1/(2+n)}$.
Both very small and very large condensates recover the size dependence $D_\infty \propto 1/R$ expected from the Stokes-Einstein relation.
Taken together, this suggests that the rate of coalescence will initially be higher due to a larger effective diffusion coefficient $D_\infty$ and consequently a higher rate of collisions (Fig.~\ref{fig::diffusion_coefficient_comparison}).
With increasing condensate radii, Brownian motion will be rapidly suppressed, in particular for $\tau_\alpha < \tau_f$ (Fig.~\ref{fig::diffusion_coefficient_size}B).
The change in scaling will create a bottleneck and skew the droplet size distribution.
Thus, interestingly, this cross-over could violate the self-similarity of the condensate size distribution on long time scales, as recently seen in experiment~\cite{Laprade_2026}.

These arguments neglect the fact that moving condensates show dipole-dipole interactions, which should further accelerate coalescence and would be interesting to study in more detail.
One simple expectation is that hydrodynamic interactions, which couple the motion of condensates over large distances, resemble the motion due to an externally applied field studied in the first part of the manuscript.
This could be the case, for example, for groups of condensates whose distance is smaller than the micron-scale correlation length of chromatin flows in the nucleus~\cite{zidovska_2013}.
Based on this reasoning, one expects to recover the accelerated coalescence $\langle R\rangle \propto t$, which is consistent with studies on the late stages of phase separation~\cite{siggia_1979}.

\section{Discussion}

\subsection{Directed coalescence by translation-induced polarization}
A core result of this manuscript is that the motion of conserved phase-separated domains with different compositions and physical properties will generically induce long-ranged forces in the form of chemical potential gradients reminiscent of electric dipole fields.
This is a universal consequence of mass conservation and therefore applies not just to condensates, but to any conserved domain-forming system.
Future research is needed to confirm the predicted phenomenon of directed coalescence, to quantitatively measure the underlying forces, and to understand how the emerging condensate-condensate interactions are affected by viscoelastic stress propagation and impermeable walls.

The theoretical predictions could be tested experimentally on a dispersion of condensates with differing sizes subject to fluid flow, gravitational deposition in an \emph{in vitro} setting, phoresis, or even optical tweezers.
Directed coalescence due to translation-induced polarization predicts that the average condensate size should grow as $\langle R \rangle \sim t^1$, which can be tested by measuring the condensate size distribution.
Moreover, coalescence by translation-induced polarization predicts signature front-seeking behavior, in which smaller condensates circle around towards the front of larger condensates, which can be tested by tracking the centroids of condensates in experiment.
Finally, the theory predicts that condensate responses are non-reciprocal according to their size, which can also be tested in experiment.

\subsection{Potential consequences for other systems}
Beyond applying external fields to control condensate dynamics, condensate motion can also emerge as a consequence of chemical reactions.
Such chemical reactions can lead to effective (screened) monopole interactions~\cite{li_2020, kumar_2023, banani_2024}, whereas the motion of condensates leads to dipole interactions, and capillary waves correspond to multipole couplings.
By combining these mechanisms, I hypothesize that chemically active, self-propelling condensates~\cite{decayeux_2021, demarchi_2023} can form stable flocks through a competition between directed coalescence due to translation-induced polarization and droplet divisions due to chemical reactions.
These ideas can find potential application to mass-conserved reaction-diffusion systems if one can map them to a generalized Cahn-Hilliard model~\cite{robinson_2024}.
This has potentially far-reaching implications for understanding the formation of dynamical patterns in reaction-diffusion systems.

\subsection{Viscoelastic stirring modifies the rate and size-dependence of Brownian motion}
This manuscript presented a framework for studying the Brownian motion of biomolecular condensates in actively driven viscoelastic media.
Active stress fluctuations were introduced phenomenologically by breaking the fluctuation-dissipation theorem.
In doing so, one discovers that, depending on the ratio between the correlation time $\tau_\alpha$ of the active stresses and the viscoelastic relaxation time $\tau_f$ of the fluid, condensate motion can be enhanced ($\tau_\alpha < \tau_f$) or suppressed ($\tau_\alpha > \tau_f$) compared to an equilibrium-like system ($\tau_\alpha = \tau_f$).
Moreover, the theory predicted how condensate motion can be decoupled from condensate size for $\tau_\alpha \gg \tau_f$, thus lifting a fundamental limitation on the frequency of collisions in polydisperse solutions.
Interestingly, recent experiments have demonstrated that condensates in self-stirred solutions can indeed show size-independent diffusion~\cite{Laprade_2026}, which supports the theoretical framework developed here.
This could potentially find applications in the development of droplet-based chemical reactors, where frequent collisions could lead to the synchronization of reactions via the exchange of reactants.
Future research will explore this possibility by taking into account enzymatic chemistry.

It would also be interesting to compare the predictions of the analytical theory with already published experimental data.
For example, recent work has shown that cytoskeletal activity quantitatively increases the frequency of collisions between biologically relevant condensates in the cell nucleus, such as speckles involved in RNA splicing~\cite{aljord_2022}.
This effect was shown to accelerate the coalescence of condensates and affect the chemical reactions within them~\cite{aljord_2022}.
To further interrogate this system, one could tune the correlation time $\tau_\alpha$ of the active stress fluctuations by changing the residence time of myosin motors on actin filaments, or the relaxation time $\tau_f$ of the viscoelastic fluid by tuning the turnover of actin or actin crosslinkers (cytoplasm).
Perturbing active loop extrusion (nucleoplasm) will likely affect both parameters.
In addition, tracking the mean squared displacement and size distribution of condensates over time would yield complementary information that would allow inference of model parameters.
This would allow testing whether cells typically operate in the $\tau_\alpha > \tau_f$ or in the $\tau_\alpha < \tau_f$ regime and open new avenues for controlling intracellular organization.

\begin{figure}[t]
    \centering
	\includegraphics{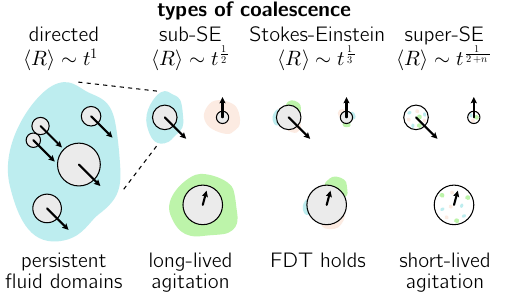}
	\caption{%
    Different types of coalescence predicted by the theory of condensates in driven media.
    If condensates are smaller and closer together than the correlated fluid domains, the theory predicts deterministic, directed coalescence with $\langle R \rangle \sim t^{1}$ for the condensate radii.
    If the time scales of agitations and fluid relaxation match so that the fluctuation-dissipation theorem holds, then diffusion follows the Stokes-Einstein relation leading to the well-known $\langle R \rangle \sim t^{1/3}$ scaling for the condensate radius.
    If agitations are long-lived and condensates are small, then diffusion is size-independent (sub-Stokes-Einstein) leading to faster scaling of the condensate radii despite lower diffusion coefficient.
    If agitations are short-lived and condensates are large, then diffusion has increased size-dependence (super-Stokes-Einstein) leading to slower scaling of the condensate radii despite higher diffusion coefficient.
	}
	\label{fig::coalescence_types}
\end{figure}

\subsection{Implications for the cell nucleoplasm and cytoplasm}
In the context of the cell nucleus, I hypothesize that both mechanisms discussed in this manuscript, namely translation-induced polarization and driven Brownian motion, can contribute to condensate coalescence [Fig.~\ref{fig::coalescence_types}].
Prior seminal work has shown that the viscoelastic chromatin in the cell nucleus undergoes correlated flows on length scales of microns (stirring length) and timescales of \SI{10}{\second}~\cite{zidovska_2013}.
The theory predicts that, relative to a reference system that satisfies the fluctuation-dissipation theorem ($\tau_\alpha = \tau_f$), active fluctuations that decorrelate quickly ($\tau_\alpha < \tau_f$) will enhance the Brownian motion of condensates whereas active fluctuations that decorrelate slowly ($\tau_\alpha > \tau_f$) will actually suppress the Brownian motion of condensates.
In absence of such a reference system, for a scenario with long-lived active fluctuations ($\tau_\alpha > \tau_f$), the Brownian motion on long time scales $D_\infty$ will appear dramatically amplified compared to the Brownian motion on short time scales $D_0$.
Moreover, in this case the condensates will show a cross-over from ballistic motion on short time scales to diffusive motion on long time scales [Fig.~\ref{fig::effective_diffusion_coefficient_appendix}].
Both this cross-over from ballistic to diffusive motion, and the size-independent diffusion coefficient predicted for $\tau_\alpha > \tau_f$ [Fig.~\ref{fig::diffusion_size_scaling}] are consistent with reconstitution experiments~\cite{Laprade_2026}.

Whether viscoelastic stirring actually enhances transport ($\tau_\alpha < \tau_f$) or only looks like it does ($\tau_\alpha > \tau_f$), it should intuitively also increase the encounter rates of the condensates.
However, viscoelastic stirring cannot bridge the final gap once condensates are closer than the stirring length corresponding to a micron-scale domain of coherently moving chromatin.
This is because below the stirring length the motion of the condensates will become correlated, and such correlations cancel out in the fluctuations of the pairwise distance.
A potential solution is that differently-sized condensates, once they are within the same micron-scale domain of coherently moving chromatin, will undergo directed coalescence due to translation-induced polarization.
This could bring together condensates at different regulatory elements, similar to how correlated fluctuations can bring together chromatin loci by tension propagation along the chromatin polymer~\cite{goychuk_2023_chromatin, goychuk_2024_chromatin}, but instead mediated by phase separation.
Biomolecular specificity could be provided by using a species of ``arbiter'' molecules which partition to both condensates and communicate with each other via translation-induced polarization.

On a final note, if material turnover is present, then the theory predicts that translation-induced polarization will have a screening length and will, moreover, affect the chemical reactions within the condensates.
If the turnover is fast enough, then condensates will mainly interact with other condensates and the majority of them will not sense any impermeable domain boundaries.
However, if material turnover is slow, then moving condensates will act akin to dipoles, and the phenomenology of dipolar media can depend on the shape of the surrounding impermeable boundary even in thermodynamic systems~\cite{Banerjee_1998}.
Although I have here discussed a nonequilibrium system, this suggests that fluctuations of impermeable domain boundaries such as the nuclear membrane or the cell membrane could potentially induce interactions among and motion of condensates without the need for bulk fluid flows.
This idea is supported by a change in perspective: relative to a fluctuating membrane as a reference point, condensates will move, and this motion will induce chemical potential gradients.

\begin{acknowledgments}
The author thanks Yannick Azhri Din Omar, Saeed Mahdisoltani, Mehran Kardar, and Arup K. Chakraborty for insightful discussions and critical reading of the manuscript.
The author also thanks Eric R. Dufresne for pointing out the possibility of directed coalescence in experiment.
This work was supported by the National Science Foundation through the Biophysics of Nuclear Condensates grant (MCB-2044895).
During his time at MIT, A.G. was supported by an EMBO Postdoctoral Fellowship (ALTF 259-2022).
Part of this work was carried out at HZI with support by the Ministry of Science and Culture of Lower Saxony through funds from the program zukunft.niedersachsen of the Volkswagen Foundation for the ‘CAIMed – Lower Saxony Center for
Artificial Intelligence and Causal Methods in Medicine’ project.
\end{acknowledgments}

%\null\cleardoublepage

%TC:ignore
\appendix

\renewcommand{\thefigure}{S\arabic{figure}}
\setcounter{figure}{0}

\section*{Technical calculations}

\section{Feedback of condensate motion on the fluid}
\label{sec:flow_interdependent}

To understand the response of the embedding fluid to condensate motion and when this feedback can be neglected, consider a generalized linear response theory for the fluid flow,
\begin{equation}
    \vec{v}_f(\vec{x},t) 
    = \int d^d\vec{x}' \int_{-\infty}^{t} dt' \,
    \mathcal{G}(\vec{x} - \vec{x}', t-t')
    \cdot \vec{f}(\vec{x}',t') \, ,
\end{equation}
with driving force $\vec{f} \coloneqq \vec\nabla \cdot \mat\alpha - c \vec\nabla\mu$.
The idea is to split the flow into two contributions, $\vec{v}_f = \vec{v}_f^\star + \delta \vec{v}_f$, where
\begin{equation}
    \vec{v}_f^\star(\vec{x},t) = \int d^d\vec{x}' \int_{-\infty}^{t} \! \! dt' \,
    \mathcal{G}(\vec{x} - \vec{x}', t-t')
    \cdot \vec\nabla \cdot \mat\alpha(\vec{x}',t')
\end{equation}
is the condensate-independent part of the flow due to active and passive forces.
The perturbation of the fluid flow by the condensates is given by
\begin{equation}
    \delta \vec{v}_f(\vec{x},t) = - \int d^d\vec{x}' \int_{-\infty}^{t} \! \! dt' \,
    \mathcal{G}(\vec{x} - \vec{x}', t-t')
    \cdot c \vec\nabla\mu(\vec{x}',t') \, .
\end{equation}
In the following, the latter will be analyzed in detail.

Using $c(\vec{x},t) = c_- + \Delta c$ for $\vec{x} \in \mathcal{D}(t)$, else $c_-$, to split the domain integral leads to
\begin{multline}
    \delta \vec{v}_f(\vec{x},t) = - \int_{-\infty}^{t} \! \! dt' \Biggl[ \\
    + c_- \int d^d\vec{x}'  \,
    \mathcal{G}(\vec{x} - \vec{x}', t-t')
    \cdot \vec\nabla\mu(\vec{x}',t') \\
    + \Delta c \int_{\mathcal{D}(t')} d^d\vec{x}'  \,
    \mathcal{G}(\vec{x} - \vec{x}', t-t')
    \cdot \vec\nabla\mu(\vec{x}',t')
    \Biggr] \, .
\end{multline}
The second line can be eliminated via integration by parts, by using fluid incompressibility $\vec\nabla\cdot\mathcal{G} = 0$ and using the fact that the Green's function $\mathcal{G}$ decays to zero in the far field.
Hence, the perturbation of the fluid flow is simplified to
\begin{multline}
\label{eq:fluid_velocity_correction}
    \delta \vec{v}_f(\vec{x},t) = - \Delta c \int_{-\infty}^{t} \! \! dt' \color{lightgray}\times\color{black} \\ \color{lightgray}\times\color{black}
    \int_{\mathcal{D}(t')} d^d\vec{x}'  \,
    \mathcal{G}(\vec{x} - \vec{x}', t-t')
    \cdot \vec\nabla\mu(\vec{x}',t') \, .
\end{multline}
The next step is to substitute the expression for the gradient of the chemical potential, Eq.~\eqref{eq:long_range_force_comoving}, with nonpotential flux $\vec{j} \coloneqq c \, \vec{v}_f$.

As before, Eq.~\eqref{eq:long_range_force_comoving} can be simplified by splitting the domain integral according to  $c(\vec{x},t) = c_- + \Delta c$ for $\vec{x} \in \mathcal{D}(t)$, else $c_-$.
The term proportional to $c_-$ is eliminated via integration by parts, after replacing $\vec\nabla_{\vec{x}} \mathcal{L}(\vec{x} - \vec{x}') = -\vec\nabla_{\vec{x}'} \mathcal{L}(\vec{x} - \vec{x}')$, by using fluid incompressibility and the fact that the fundamental solution $\mathcal{L}$ of the Poisson equation decays to zero in the far field.
Taken together, the chemical potential profile in the laboratory frame is given by:
\begin{multline}
\label{eq:fluid_motion_chemical_potential}
    \vec\nabla\mu(\vec{x},t) = 
    - \vec\nabla\Psi(\vec{x},t) 
    + \frac{\Delta c}{M} \, \mathcal{B}(\vec{x} - \vec{r}) \cdot \vec{v}(t) \\
    - \frac{\Delta c}{M} \, \int_{\mathcal{D}(t)} d^d\vec{x}' \, \left[
    \vec\nabla \otimes \vec\nabla \mathcal{L}(\vec{x} - \vec{x}')
    \right] \cdot \vec{v}_f(\vec{x}',t)
    \, ,
\end{multline}
where $\vec{r} \equiv \vec{r}(t)$ is the centroid of the condensate.
The second term demonstrates how the motion of the condensate, with velocity $\vec{v}$, induces a long-range force on the fluid, with the Green's tensor $\mathcal{B}$ for an electrostatic dipole field defined by Eq.~\eqref{eq:tensor_longranged_force}.
Together with Eq.~\eqref{eq:fluid_velocity_correction}, this result illuminates that the effect of condensate motion on fluid flow is $\mathcal{O}(\Delta c^2)$ and that it decays with the mobility of the phase-separating molecules, $\propto 1/M$.
The present manuscript neglects these higher-order corrections to the fluid flow.

Nevertheless, I will briefly comment on how to make such corrections for spherical condensates with radius $R$, center of mass position $\vec{r}(t) \in \mathcal{D}(t)$ and velocity $\vec{v} \equiv \vec{v}(t)$.
This will be illustrated for a single condensate and is straightforward to generalize to a collection of topologically distinct phase-separated domains.
In essence, after inserting the chemical potential gradient generated by the condensate, Eq.~\eqref{eq:fluid_motion_chemical_potential}, into Eq.~\eqref{eq:fluid_velocity_correction}, one has:
\begin{multline}
    \delta \vec{v}_f(\vec{x},t) = \frac{\Delta c^2}{M} \int_{-\infty}^{t} \! \! dt' 
    \int_{\mathcal{D}(t')} d^d\vec{x}'
    \Biggl\{ \\
    + \mathcal{G}(\vec{x} - \vec{x}', t-t')
    \cdot 
    \Biggl[
    \frac{\vec{v}(t')}{d}
    + \frac{M\vec\nabla\Psi(\vec{x}',t')}{\Delta c} \Biggr] \\
    + \mathcal{G}'(\vec{x} - \vec{r}', \vec{x}' - \vec{r}', t-t') \cdot \vec{v}_f(\vec{x}',t') 
    \Biggr\}
    \, ,
\end{multline}
where the time-dependent domain $\mathcal{D}(t')$ of the condensate is centered around $\vec{r}' \equiv \vec{r}(t')$ and $\mathcal{B}(\vec{x}' - \vec{r}') = - \mat{I}/d$ for $|\vec{x}' - \vec{r}'| \leq R$.
The second line shows that the fluid responds to lowest order as if driven by a point force, although the force generated by condensate motion is nonlocal.

Because $\vec{v}_f = \vec{v}_f^\star + \delta \vec{v}_f$, the third line indicates that the disturbance $\delta \vec{v}_f$ of the fluid flow must be calculated self-consistently.
Here, 
\begin{equation}
    \mathcal{G}'(\vec{x}, \vec{x}', t) \coloneqq \int_{\mathcal{D}} \! d^d\vec{z} \,
    \mathcal{G}(\vec{x} - \vec{z}, t)
    \cdot 
    \vec\nabla \otimes \vec\nabla \mathcal{L}(\vec{z} - \vec{x}') \, ,
\end{equation}
with $\mathcal{D}$ the domain of the condensate in the comoving frame, is a modified Green's function.
In future studies, these results can be used to derive an equivalent of the Rotne-Prager-Yamakawa tensor~\cite{rotne_prager_1969, yamakawa_1970} to study fluid-mediated interactions in the context of biomolecular condensates.

\section{Details of series expansion of the fluid flow correlation function}
\label{sec:flow_expansion}

Starting from the Taylor series expansion of the fluid flow around the trajectory midpoint, Eq.~\eqref{eq:flow_expansion_instantaneous} in the main text, I will now detail the algebraic steps to arrive at Eq.~\eqref{eq:flow_expansion}.
Based on this expansion, the covariance of the fluid velocity is given by:
\begin{multline}
    \langle\vec{v}_f(\vec{z}+\vec{r},t) \cdot \vec{v}_f(\vec{z}'+\vec{r}',t')\rangle 
    = \sum_{nm} \frac{1}{n! m!} 
    \Bigl\langle 
    \left[\tfrac{\vec{r}  - \vec{r}'}{2} \cdot \vec\nabla_{\vec{z}}\right]^n \color{lightgray} \times \color{black} \\ \color{lightgray} \times \color{black}
    \left[\tfrac{\vec{r}' - \vec{r}}{2} \cdot \vec\nabla_{\vec{z}'}\right]^m 
    \vec{v}_f(\vec{z} + \bar{\vec{r}}, t) \cdot \vec{v}_f(\vec{z}' + \bar{\vec{r}}, t') 
    \Bigr\rangle
    \, ,
\end{multline}
where I have substituted the midpoint $\bar{\vec{r}} \coloneqq (\vec{r} + \vec{r}')/2$ of the condensate trajectory in the time window $[t,t']$.
The key idea is to now split the correlations of the condensate position and the correlations of the fluid flow.
As discussed in the main text, to that end I make the preaveraging approximation $\vec{v}_f(\vec{z} + \bar{\vec{r}},t) \cdot \vec{v}_f(\vec{z}' + \bar{\vec{r}},t') \approx \langle \vec{v}_f(\vec{z} + \bar{\vec{r}},t) \cdot \vec{v}_f(\vec{z}' + \bar{\vec{r}},t') \rangle$, which essentially posits that the fluid fluctuates much faster than the condensate moves.
Together with substituting the definition of the covariance of the hydrodynamic fluctuations, $\mathcal{C}_f(\vec{z} - \vec{z}', t - t') \coloneqq\langle\vec{v}_f(\vec{z}, t) \cdot \vec{v}_f(\vec{z}', t') \rangle$, and using $\vec\nabla_{\vec{z}'}\mathcal{C}_f(\vec{z} - \vec{z}', t - t') = -\vec\nabla_{\vec{z}}\mathcal{C}_f(\vec{z} - \vec{z}', t - t')$, this yields
\begin{multline}
\label{eq:correlation_expansion_intermediate}
    \langle\vec{v}_f(\vec{z}+\vec{r},t) \cdot \vec{v}_f(\vec{z}'+\vec{r}',t')\rangle 
    \approx \sum_{nm} \frac{1}{n! m!} \color{lightgray} \times \color{black} \\ \color{lightgray} \times \color{black}
    \left\langle 
    \left[\frac{\vec{r}  - \vec{r}'}{2} \cdot \vec\nabla_{\vec{z}}\right]^{n+m}   
    \right\rangle \,
    \mathcal{C}_f(\vec{z} - \vec{z}', t - t')
    \, .
\end{multline}
After using Wick's theorem, one has
\begin{multline}
    \langle\vec{v}_f(\vec{z}+\vec{r},t) \cdot \vec{v}_f(\vec{z}'+\vec{r}',t')\rangle 
    \approx \sum_{\substack{nm \\ n+m \text{ even}}} \frac{1}{n! m!} \frac{(n+m)!}{\left(\frac{n+m}{2}\right)!}
    \color{lightgray} \times \color{black} \\ \color{lightgray} \times \color{black}
    \frac{1}{2^{\frac{n+m}{2}}}
    \left\langle 
    \left[\frac{\vec{r}  - \vec{r}'}{2} \cdot \vec\nabla_{\vec{z}}\right]^{2}   
    \right\rangle^{\frac{n+m}{2}} \,
    \mathcal{C}_f(\vec{z} - \vec{z}', t - t')
    \, .
\end{multline}
This expression can be simplified by reordering the summation using the integer $l = \frac{n+m}{2}$ and by exploiting isotropy, $\langle \Delta\vec{r} \otimes \Delta\vec{r} \rangle = \langle |\Delta\vec{r}|^2\rangle \, \mat{I}/d$, leading to:
\begin{multline}
    \langle\vec{v}_f(\vec{z}+\vec{r},t) \cdot \vec{v}_f(\vec{z}'+\vec{r}',t')\rangle 
    \approx 
    \sum_l \sum_{n=0}^{2l}
    \frac{(2l)!}{n! (2l - n)!} 
    \color{lightgray} \times \color{black} \\ \color{lightgray} \times \color{black}
    \frac{1}{l!}
    \frac{1}{2^l}
    \left[
    \frac{1}{4d}
    \left\langle 
    [\vec{r}  - \vec{r}']^2
    \right\rangle  
    \vec\nabla_{\vec{z}}^2
    \right]^l \,
    \mathcal{C}_f(\vec{z} - \vec{z}', t - t')
    \, .
\end{multline}
Finally, after evaluating the inner sum, one arrives at the Taylor series of Eq.~\eqref{eq:flow_expansion} reported in the main text:
\begin{multline}
    \langle\vec{v}_f(\vec{z}+\vec{r},t) \cdot \vec{v}_f(\vec{z}'+\vec{r}',t')\rangle \\
    \approx 
    \sum_l 
    \frac{1}{l!}
    \left[
    \frac{1}{2d}
    \left\langle 
    [\vec{r}  - \vec{r}']^2
    \right\rangle  
    \vec\nabla_{\vec{z}}^2
    \right]^l \,
    \mathcal{C}_f(\vec{z} - \vec{z}', t - t') \\
    = \exp\biggl[
    \frac{1}{2d}
    \left\langle 
    [\vec{r}  - \vec{r}']^2
    \right\rangle  
    \vec\nabla_{\vec{z}}^2
    \biggr] \, \mathcal{C}_f(\vec{z} - \vec{z}', t - t')
    \, .
\tag{\ref{eq:flow_expansion}}
\end{multline}

\section{Differential equation for mean squared displacement in Fourier space representation}
\label{sec:velocity_autocorrelation_equation}

For spherical condensates with radius $R$, the differential equation Eq.~\eqref{eq:ode_transformed_final} can be cast into a numerically more convenient form via a Fourier transform in space and time,
\begin{align}
    \widehat{\mathcal{C}}_f(|\vec{q}|, \omega) &\coloneqq \int d^d\vec{z} \int dt \, e^{-i\vec{q}\cdot\vec{z} - i\omega t} \, \mathcal{C}_f(\vec{z}, t) \, , \\
    \mathcal{C}_f(\vec{z},t) &\coloneqq \int \frac{d^d\vec{q}}{(2\pi)^d} \int \frac{d\omega}{2\pi} \, e^{i\vec{q}\cdot\vec{z} + i \omega t} \, \widehat{\mathcal{C}}_f(|\vec{q}|, \omega) \, ,
\end{align}
where I have taken into account the isotropy of the correlation function.
Substitution in Eq.~\eqref{eq:ode_transformed_final} leads to
\begin{multline}
\label{eq:ode_transformed}
    \frac{1}{2} \partial_t^2  
    \MSD(t)  
    = D_c \, \delta(t) 
    + \frac{d \,\Gamma\bigl(1+\tfrac{d}{2}\bigr)}{\pi^{d/2} R^d}
    \int \frac{d\omega}{2\pi} e^{i \omega t} \, 
    \color{lightgray}\times\color{black} \\ \color{lightgray}\times\color{black}
    \int_0^{\infty} \! \frac{dq}{q} \,  
    \exp\biggl[
    -\frac{q^2}{2d}
    \MSD(t)
    \biggr] \, 
    J_{d/2}^2(q R) \, 
    \widehat{\mathcal{C}}_f(q, \omega) 
    \, ,
\end{multline}
with the form factor given by the Bessel function of the first kind,
\begin{equation}
    J_{d/2}(q)
    = \frac{q^{d/2}}{2^{d/2} \Gamma\bigl(1+\tfrac{d}{2}\bigr)} \int_{|\vec{x}| \leq 1} \frac{d^d\vec{x}}{V_1} e^{i\vec{q}\cdot \vec{x}} \, ,
\end{equation}
where $V_1$ is the volume of the 1-sphere.

If the high-frequency modes of the fluid flow do not decay to zero,
\begin{equation}
    \widehat{\mathcal{C}}_f(q, \infty) \coloneqq
    \lim_{\omega\to\infty}\widehat{\mathcal{C}}_f(q, \omega) \neq 0 \, ,
\end{equation}
then this will lead to an additional $\delta$-correlated contribution in Eq.~\eqref{eq:ode_transformed_final}.
To manage this contribution, I split the correlation function of the fluid flow, 
\begin{equation}
\label{eq:split_correlation_function}
    \widehat{\mathcal{C}}_f(q, \omega) = \widehat{\mathcal{C}}_f(q, \infty) + \left[\widehat{\mathcal{C}}_f(q, \omega) - \widehat{\mathcal{C}}_f(q, \infty)\right] \, ,
\end{equation}
and use the trivial initial condition $\MSD(0) = 0$ to rewrite the Fourier-transformed dynamics, Eq.~\eqref{eq:ode_transformed}:
\begin{multline}
\label{eq:ode_transformed_split}
    \frac{1}{2} \partial_t^2  
    \MSD(t)  
    = D_c \, \delta(t) \\
    + \frac{d \,\Gamma\bigl(1+\tfrac{d}{2}\bigr)}{\pi^{d/2} R^d} 
    \int_0^{\infty} \! \frac{dq}{q} \,   
    J_{d/2}^2(q R) \, 
    \widehat{\mathcal{C}}_f(q, \infty) \,
    \delta(t)
    \\
    + \frac{d \,\Gamma\bigl(1+\tfrac{d}{2}\bigr)}{\pi^{d/2} R^d}
    \int \frac{d\omega}{2\pi} e^{i \omega t} \,
    \int_0^{\infty} \! \frac{dq}{q}  
    \exp\biggl[
    -\frac{q^2}{2d}
    \MSD(t)
    \biggr] 
    \color{lightgray}\times\color{black} \\ \color{lightgray}\times\color{black}  
    J_{d/2}^2(q R) \, 
    \left[\widehat{\mathcal{C}}_f(q, \omega) - \widehat{\mathcal{C}}_f(q, \infty)\right] 
    \, .
\end{multline}
Note that the mean squared displacement is a symmetric function under time reversal.
Integrating Eq.~\eqref{eq:ode_transformed_split} within an infinitesimally small region around the origin, yields the second initial condition, $\partial_t \MSD(0^{+}) = D_0$, where
\begin{equation}
    D_0 = D_c 
    + \frac{d \,\Gamma\bigl(1+\tfrac{d}{2}\bigr)}{\pi^{d/2} R^d} 
    \int_0^{\infty} \! \frac{dq}{q} \,   
    J_{d/2}^2(q R) \, 
    \widehat{\mathcal{C}}_f(q, \infty) \, .
\tag{\ref{eq:ode_transformed_boundary_condition}}
\end{equation}
For $t>0$, where the $\delta$-distributions vanish, Eq.~\eqref{eq:ode_transformed_split} reduces to
\begin{multline}
\label{eq:ode_transformed_final_fourierspace}
    \frac{1}{2} \partial_t^2  
    \MSD(t)  
    = 
    \frac{d \,\Gamma\bigl(1+\tfrac{d}{2}\bigr)}{\pi^{d/2} R^d}
    \int_0^{\infty} \! \frac{dq}{q} 
    \exp\biggl[
    -\frac{q^2}{2d}
    \MSD(t)
    \biggr]
    \color{lightgray}\times\color{black} \\ \color{lightgray}\times\color{black}  
    J_{d/2}^2(q R) \, 
    \int\!\frac{d\omega}{2\pi} e^{i \omega t}
    \left[
    \widehat{\mathcal{C}}_f(q, \omega) - \widehat{\mathcal{C}}_f(q, \infty)
    \right] 
    \, .
\end{multline}
Since the $\delta$-distributions do not matter for $t>0$ and I have lifted them in Eq.~\eqref{eq:ode_transformed_final_fourierspace} by deriving appropriate initial conditions, one can also drop the corresponding terms in Eq.~\eqref{eq:ode_transformed_final}.

\section{Derivation of asymptotic dynamics on short and long time scales}
\label{sec:asymptotics}

The dynamics on short time scales is dominated by the initial condition on the slope of the mean squared displacement, $\partial_t \MSD(0^{+}) = D_0$, with the effective diffusion coefficient given by Eq.~\eqref{eq:ode_transformed_boundary_condition}.
Hence, the mean squared displacement grows linearly with time initially, $\MSD(t) \approx D_0 |t|$.
To understand the dynamics on long time scales, it is useful to integrate Eq.~\eqref{eq:ode_transformed} forward in time:
\begin{multline}
    \partial_t  
    \MSD(t)  
    = 
    2\int_0^t dt'\Biggl[ 
    D_c \, \delta(t') 
    + \frac{d \,\Gamma\bigl(1+\tfrac{d}{2}\bigr)}{\pi^{d/2} R^d}
    \int \frac{d\omega}{2\pi} e^{i \omega t'} 
    \color{lightgray}\times\color{black} \\ \color{lightgray}\times\color{black}
    \int_0^{\infty} \! \frac{dq}{q} \,  
    \exp\biggl[
    -\frac{q^2}{2d}
    \MSD(t')
    \biggr] \, 
    J_{d/2}^2(q R) \, 
    \widehat{\mathcal{C}}_f(q, \omega) 
    \Biggr]
    \, .
\end{multline}
This expression can be further manipulated by symmetrizing the integration domain, which is enabled by the time-reversal symmetry of the mean squared displacement.
Moreover, note that the exponential in the second line has its peak at $t' = 0$ and then monotonically decays.
Similar to a saddle-point approximation, one can assume that this decay is dominated by the dynamics on short time scales, where $\MSD(t) \approx D_0 |t|$.
This approximation breaks down if $\partial_t \MSD(t)$ starts at very small values and then increases dramatically.

Hence, the growth of the mean squared displacement on large time scales is governed by another effective diffusion coefficient:
\begin{multline}
    D_\infty  
    \approx 
    D_c
    + \int_{-\infty}^\infty dt'  
     \frac{d \,\Gamma\bigl(1+\tfrac{d}{2}\bigr)}{\pi^{d/2} R^d}
    \int \frac{d\omega}{2\pi} e^{i \omega t'} 
    \color{lightgray}\times\color{black} \\ \color{lightgray}\times\color{black}
    \int_0^{\infty} \! \frac{dq}{q} \,  
    \exp\biggl[
    -\frac{q^2}{2d}
    D_0 |t'|
    \biggr] \, 
    J_{d/2}^2(q R) \, 
    \widehat{\mathcal{C}}_f(q, \omega) 
    \, ,
\end{multline}
where I have defined $D_\infty \coloneqq \lim_{t\to\infty} \partial_t \MSD(t)$.
This the Fourier space representation of Eq.~\eqref{eq:diffusion_coefficient_long} in the main text.
Splitting the correlation function of the fluid flow according to Eq.~\eqref{eq:split_correlation_function}, and carrying out the temporal Fourier transform, one finds:
\begin{multline}
    D_\infty  
    \approx D_0
    + \frac{d \,\Gamma\bigl(1+\tfrac{d}{2}\bigr)}{\pi^{d/2} R^d}
    \int_0^{\infty} \! \frac{dq}{q} \,
    J_{d/2}^2(q R) \,  
    \color{lightgray}\times\color{black} \\ \color{lightgray}\times\color{black}  
    \int \frac{d\omega}{2\pi}
    \frac{4 d D_0 q^2}{4 d^2 \omega ^2 + D_0^2 q^4}  
    \left[\widehat{\mathcal{C}}_f(q, \omega) - \widehat{\mathcal{C}}_f(q, \infty)\right]
    \, .
\end{multline}
The second term determines whether condensate motion on large time scales is faster or slower than condensate motion on short time scales.

%\section{Anomalous transport at intermediate time scales}

\section{Details of numerical simulations}
This supporting section details the numerical calculation of the mean squared displacement of spherical condensates with radius $R$ in $d=3$ dimensions.
As discussed in the main text, the first term in Eq.~\eqref{eq:dynamics_3d_short_time scale} is negligible for most practical scenarios and one can approximate $D_c \approx 0$.
In the following, $\Lambda \coloneqq \tau_f / \tau_\alpha$ is the ratio between the fluid relaxation time and the stirring time, $s \coloneqq t / \tau_\alpha$ is the nondimensionalized time and $X(t) \coloneqq \MSD(t)/R^2$ is the nondimensionalized mean squared displacement.
Taken together, with the Fourier number defined by Eq.~\eqref{eq:parameter_estimation_fourier_number}, the initial condition $\partial_t \MSD(0^{+}) = D_0$ corresponds to
\begin{equation}
    \partial_s X(0^{+}) = \text{Fo} \, \Lambda^2 \, ,
\end{equation}
and the initial condition $\text{MSD(0)} = 0$ corresponds to $X(0)=0$.

To obtain the mean squared displacement of the condensate, one has to solve Eq.~\eqref{eq:ode_transformed_final}, which is more convenient after representing the spatial integrals in Fourier space, Eq.~\eqref{eq:ode_transformed_final_fourierspace}.
For the correlation of the fluid flows, I used Eq.~\eqref{eq:fluid_correlation} with negligible hydrodynamic screening, $\gamma \approx 0$, for simplicity.
In summary, the results reported in this manuscript were obtained by solving the following nondimensionalized differential equation:
\begin{equation}
    \frac{1}{2} \partial_s^2  
    X 
    = 
    \frac{15}{4} \text{Fo} \, 
    \left[1 - \Lambda^2\right]
    e^{-s}
    K(X)
    \, ,
\end{equation}
where
\begin{equation}
    K(X) \coloneqq 
    \int_0^{\infty} \! \frac{dq}{q^3} 
    \exp\biggl[
    -\frac{q^2}{6}
    X
    \biggr]  
    J_{3/2}^2(q) \, .
\end{equation}

The asymptotic dynamics on short time scales are given by [Eq.~\eqref{eq:ode_transformed_boundary_condition}]
\begin{equation}
    \frac{D_0 \tau_\alpha}{R^2} \approx \text{Fo} \, \Lambda^2 
    \, ,
\end{equation}
and on long time scales are given by [Eq.~\eqref{eq:diffusion_coefficient_long}]
\begin{multline}
    \frac{D_\infty \tau_\alpha}{R^2}  
    \approx \text{Fo} \, \Lambda^2 \Biggl[ 
    1
    + 
    45 \frac{1 - \Lambda^2}{\Lambda^2} \,
    \int_0^{\infty} \! \frac{dq}{q^3} \,
    \frac{J_{3/2}^2(q)}{6 + \text{Fo} \, \Lambda^2 q^2} 
    \Biggr]
    \, .
\end{multline}
%
%Interestingly, this suggests that the diffusivity of phase-separated domains is linear in the Fourier number.
%This is different from the diffusion of individual tracer particles in regular fluids under shear flow, where one would expect a quadratic dependence on the Fourier number.
%However, a linear dependence has been experimentally observed for tracer particles in supercooled liquids and modeled as an escape problem in a random potential landscape under shear flow~\cite{zaccone_2025}.

%TC:endignore

\cleardoublepage

\bibliography{bibliography}

@article{goychuk_2023_chromatin, 
title={Polymer folding through active processes recreates features of genome organization}, 
volume={120}, 
ISSN={0027-8424}, 
url={https://doi.org/10.1073/pnas.2221726120}, 
DOI={10.1073/pnas.2221726120}, 
number={20}, 
journal={Proc. Natl. Acad. Sci. U.S.A.}, 
author={Goychuk, Andriy and Kannan, Deepti and Chakraborty, Arup K. and Kardar, Mehran}, 
year={2023}, 
month=may, 
pages={e2221726120} 
}

@article{goychuk_2024_chromatin, 
title={Delayed Excitations Induce Polymer Looping and Coherent Motion}, 
volume={133}, 
ISSN={0031-9007}, 
DOI={10.1103/physrevlett.133.078101}, 
number={7}, 
journal={Phys. Rev. Lett.}, 
author={Goychuk, Andriy and Kannan, Deepti and Kardar, Mehran}, 
year={2024}, 
pages={078101} 
}

@article{Banerjee_1998, 
title={Thermodynamic Limit for Dipolar Media}, 
volume={93}, 
ISSN={0022-4715}, 
DOI={10.1023/b:joss.0000026729.83187.79}, 
number={1–2}, 
journal={J. Stat. Phys.}, 
author={Banerjee, S and Griffiths, R B and Widom, M}, 
year={1998}, 
pages={109–141} 
}

@article{Laprade_2026, 
title={The coarsening of biomimetic condensates in an active fluid is non-self-similar}, 
ISSN={1745-2473}, 
DOI={10.1038/s41567-026-03191-w}, 
journal={Nat. Phys.}, 
author={Laprade, Jeremy and Frechette, Layne B. and Amey, Christopher and Cusi, Adrielle T. and Baskaran, Aparna and Rogers, W. Benjamin and Duclos, Guillaume}, 
year={2026}, 
pages={1–8} 
}

@article{batchelor_1976, 
title={Brownian diffusion of particles with hydrodynamic interaction}, 
volume={74}, 
ISSN={0022-1120}, 
url={https://www.cambridge.org/core/product/8A2AB2E1765EA4CE8696C27C33A951D2}, DOI={10.1017/s0022112076001663}, 
number={1}, 
journal={J. Fluid Mech.}, 
author={Batchelor, G K}, 
year={1976}, 
pages={1–29} 
}

@article{bouchaud_1990, 
year = {1990}, 
title = {{Anomalous diffusion in disordered media: Statistical mechanisms, models and physical applications}}, 
author = {Bouchaud, Jean-Philippe and Georges, Antoine}, 
journal = {Phys. Rep.}, 
issn = {0370-1573}, 
doi = {10.1016/0370-1573(90)90099-n}, 
url = {https://www.sciencedirect.com/science/article/pii/037015739090099N},  
pages = {127--293}, 
number = {4-5}, 
volume = {195}, 
month = {11}, 
}

@article{romano_2025preprint,
author = {Romano, Jacopo and Golestanian, Ramin and Mahault, Benoît},
title  = {Dynamics of phase-separated interfaces in inhomogeneous and driven mixtures},
journal  = {Soft Matter},
year  = {2025},
volume  = {21},
issue  ={48},
pages  ={9245-9256},
publisher  ={The Royal Society of Chemistry},
doi  ={10.1039/D5SM00625B},
url  ={http://dx.doi.org/10.1039/D5SM00625B},
}

@article{debye_huckel_1923, 
year = {1923}, 
title = {{Zur Theorie der Elektrolyte. I. Gefrierpunktserniedrigung und verwandte Erscheinungen}}, 
author = {H\"uckel, E and Debye, P}, 
journal = {Phys. Z}, 
pages = {185--206}, 
volume = {24}, 
}

@article{zwicker_2022, 
year = {2022}, 
title = {{The intertwined physics of active chemical reactions and phase separation}}, 
author = {Zwicker, David}, 
journal = {Curr. Opin. Colloid Interface Sci.}, 
issn = {1359-0294}, 
doi = {10.1016/j.cocis.2022.101606}, 
eprint = {2202.13646},  
pages = {101606}, 
volume = {61}, 
}

@book{balian_2006,
author = {Balian, Roger},
publisher = {Springer, Berlin, Heidelberg},
title = {From Microphysics to Macrophysics},
year = {2007},
isbn = {3540454691},
month = nov,
ean = {9783540454694},
pagetotal = {488},
url = {https://link.springer.com/book/9783540454694},
}

@Book{degroot_2013,
author = {De Groot, S. R. and Mazur, P.},
publisher = {Dover Publications},
title = {Non-Equilibrium Thermodynamics},
year = {2013},
isbn = {9780486153506},
series = {Dover Books on Physics},
url = {https://books.google.com/books?id=HFAIv43rlGkC},
}

@article{einstein_1905, 
year = {1905}, 
title = {{Über die von der molekularkinetischen Theorie der Wärme geforderte Bewegung von in ruhenden Flüssigkeiten suspendierten Teilchen}}, 
author = {Einstein, A.}, 
journal = {Ann. Phys.}, 
issn = {0003-3804}, 
doi = {10.1002/andp.19053220806}, 
url = {https://doi.org/10.1002/andp.19053220806}, 
pages = {549--560}, 
number = {8}, 
volume = {322}, 
month = {01}
}

@book{landau_lifshitz_1987, 
year = {1987}, 
title = {{Fluid Mechanics}}, 
author = {Landau, L. D. and Lifshitz, E. M.}, 
isbn = {978-0-08-057073-0}, 
publisher = {Butterworth-Heinemann}, 
address = {Oxford}, 
edition = {Second Edition}, 
month = {00}
}

@article{bell_2025_comment, 
year = {2025}, 
title = {{Comment on ``Brownian motion of droplets induced by thermal noise''}}, 
author = {Bell, J. B. and Nonaka, A. and Garcia, A. L.}, 
journal = {Phys. Rev. E}, 
doi = {10.1103/physreve.111.056201}, 
url = {https://link.aps.org/doi/10.1103/PhysRevE.111.056201}, 
pages = {056201}, 
number = {5}, 
volume = {111}, 
month = {05}
}

@article{lee_2023, 
year = {2023}, 
title = {{Size distributions of intracellular condensates reflect competition between coalescence and nucleation}}, 
author = {Lee, Daniel S. W. and Choi, Chang-Hyun and Sanders, David W. and Beckers, Lien and Riback, Joshua A. and Brangwynne, Clifford P. and Wingreen, Ned S.}, 
journal = {Nat. Phys.}, 
issn = {1745-2473}, 
doi = {10.1038/s41567-022-01917-0}, 
pmid = {37073403}, 
pmcid = {PMC10104779}, 
url = {https://doi.org/10.1038/s41567-022-01917-0}, 
pages = {586--596}, 
number = {4}, 
volume = {19}, 
month = {04}, 
}

@article{barker_2023, 
year = {2023}, 
title = {{Fluctuating hydrodynamics and the Rayleigh–Plateau instability}}, 
author = {Barker, Bryn and Bell, John B. and Garcia, Alejandro L.}, 
journal = {Proc. Natl. Acad. Sci. U.S.A.}, 
issn = {0027-8424}, 
doi = {10.1073/pnas.2306088120}, 
pmid = {37463215}, 
pmcid = {PMC10372655}, 
url = {https://doi.org/10.1073/pnas.2306088120},  
pages = {e2306088120}, 
number = {30}, 
volume = {120}, 
note = {doi: 10.1073/pnas.2306088120}, 
month = {07}, 
}

@book{happel_brenner_1983, 
year = {1983}, 
title = {{Low Reynolds number hydrodynamics, with special applications to particulate media}}, 
author = {Happel, John and Brenner, Howard}, 
isbn = {9789024728770}, 
series = {Mech. Fluids Transp. Process.}, 
doi = {10.1007/978-94-009-8352-6}, 
}

@article{cates_review_2018, 
year = {2018}, 
title = {{Theories of binary fluid mixtures: from phase-separation kinetics to active emulsions}}, 
author = {Cates, Michael E. and Tjhung, Elsen}, 
journal = {J. Fluid Mech.}, 
issn = {0022-1120}, 
doi = {10.1017/jfm.2017.832}, 
eprint = {1806.01239}, 
url = {https://www.cambridge.org/core/product/5BD133CB20D89F47E724D77C296FEF80},  
pages = {P1}, 
volume = {836}, 
}

@article{wilken_2023, 
year = {2023}, 
title = {{Spatial Organization of Phase-Separated DNA Droplets}}, 
author = {Wilken, Sam and Chaderjian, Aria and Saleh, Omar A.}, 
journal = {Phys. Rev. X}, 
doi = {10.1103/physrevx.13.031014}, 
url = {https://link.aps.org/doi/10.1103/PhysRevX.13.031014},  
pages = {031014}, 
number = {3}, 
volume = {13}, 
month = {08}, 
}

@article{schmitt_stark_2016, 
year = {2016}, 
title = {{Active Brownian motion of emulsion droplets: Coarsening dynamics at the interface and rotational diffusion}}, 
author = {Schmitt, M. and Stark, H.}, 
journal = {The European Physical Journal E}, 
issn = {1292-8941}, 
doi = {10.1140/epje/i2016-16080-y}, 
pmid = {27562831}, 
eprint = {1608.02571}, 
url = {https://doi.org/10.1140/epje/i2016-16080-y}, 
pages = {80}, 
number = {8}, 
volume = {39}, 
month = {08}, 
}

@article{squires_2010, 
year = {2010}, 
title = {{Fluid Mechanics of Microrheology}}, 
author = {Squires, Todd M. and Mason, Thomas G.}, 
journal = {Annual Review of Fluid Mechanics}, 
issn = {0066-4189}, 
doi = {10.1146/annurev-fluid-121108-145608}, 
pages = {413--438}, 
number = {1}, 
volume = {42}, 
}

@article{tanaka_2000, 
year = {2000}, 
title = {{Viscoelastic phase separation}}, 
author = {Tanaka, Hajime}, 
journal = {Journal of Physics: Condensed Matter}, 
issn = {0953-8984}, 
doi = {10.1088/0953-8984/12/15/201}, 
url = {https://dx.doi.org/10.1088/0953-8984/12/15/201},  
pages = {R207}, 
number = {15}, 
volume = {12}, 
month = {04}, 
}

@article{frey_kroy_2005, 
year = {2005}, 
title = {{Brownian motion: a paradigm of soft matter and biological physics}}, 
author = {Frey, E and Kroy, K}, 
journal = {Annalen der Physik}, 
issn = {0003-3804}, 
doi = {10.1002/andp.200410132}, 
url = {https://doi.org/10.1002/andp.200410132}, 
pages = {20--50}, 
number = {1-3}, 
volume = {14}, 
month = {02}, 
}

@article{driscoll_2019, 
year = {2019}, 
title = {{Leveraging collective effects in externally driven colloidal suspensions: experiments and simulations}}, 
author = {Driscoll, Michelle and Delmotte, Blaise}, 
journal = {Curr. Opin. Colloid Interface Sci.}, 
issn = {1359-0294}, 
doi = {10.1016/j.cocis.2018.10.002}, 
eprint = {1810.12977}, 
url = {https://www.sciencedirect.com/science/article/pii/S1359029418300876}, 
pages = {42--57}, 
volume = {40}, 
month = {04}, 
}

@article{cahn_1961, 
year = {1961}, 
title = {{On spinodal decomposition}}, 
author = {Cahn, John W}, 
journal = {Acta Metall.}, 
issn = {0001-6160}, 
doi = {10.1016/0001-6160(61)90182-1}, 
url = {https://www.sciencedirect.com/science/article/pii/0001616061901821},  
pages = {795--801}, 
number = {9}, 
volume = {9}, 
month = {09}
}

@article{siggia_1979, 
year = {1979}, 
title = {{Late stages of spinodal decomposition in binary mixtures}}, 
author = {Siggia, Eric D.}, 
journal = {Phys. Rev. A}, 
issn = {1050-2947}, 
doi = {10.1103/physreva.20.595}, 
url = {https://link.aps.org/doi/10.1103/PhysRevA.20.595},  
pages = {595--605}, 
number = {2}, 
volume = {20}, 
month = {08}
}

@article{shimizu_2015, 
year = {2015}, 
title = {{A novel coarsening mechanism of droplets in immiscible fluid mixtures}}, 
author = {Shimizu, Ryotaro and Tanaka, Hajime}, 
journal = {Nat. Commun.}, 
issn = {2041-1723}, 
doi = {10.1038/ncomms8407}, 
pmid = {26077672}, 
eprint = {1509.03385}, 
url = {https://doi.org/10.1038/ncomms8407}, 
pages = {7407}, 
number = {1}, 
volume = {6}, 
month = {06}
}

@article{voorhees_1992, 
year = {1992}, 
title = {{Ostwald Ripening of Two-Phase Mixtures}}, 
author = {Voorhees, P W}, 
journal = {Annu. Rev. Mater. Res.}, 
issn = {0084-6600}, 
doi = {10.1146/annurev.ms.22.080192.001213}, 
pages = {197--215}, 
number = {1}, 
volume = {22}, 
}

@article{joshi_2015, 
year = {2015}, 
title = {{Computational Modeling of Multiphase Reactors}}, 
author = {Joshi, J.B. and Nandakumar, K.}, 
journal = {Annu. Rev. Chem. Biomol. Eng.}, 
issn = {1947-5438}, 
doi = {10.1146/annurev-chembioeng-061114-123229}, 
pmid = {26134737}, 
url = {https://www.annualreviews.org/content/journals/10.1146/annurev-chembioeng-061114-123229}, 
pages = {1--32}, 
number = {1}, 
volume = {6}
}

@article{bazant_2013, 
year = {2013}, 
title = {{Theory of Chemical Kinetics and Charge Transfer based on Nonequilibrium Thermodynamics}}, 
author = {Bazant, Martin Z.}, 
journal = {Acc. Chem. Res.}, 
issn = {0001-4842}, 
doi = {10.1021/ar300145c}, 
url = {https://doi.org/10.1021/ar300145c}, 
pages = {1144--1160}, 
number = {5}, 
volume = {46}, 
note = {doi: 10.1021/ar300145c}, 
month = {05}
}

@article{cahn_hilliard_1958, 
year = {1958}, 
title = {{Free Energy of a Nonuniform System. I. Interfacial Free Energy}}, 
author = {Cahn, John W and Hilliard, John E}, 
journal = {J. Chem. Phys.}, 
issn = {0021-9606}, 
doi = {10.1063/1.1744102}, 
url = {https://doi.org/10.1063/1.1744102},  
pages = {258--267}, 
number = {2}, 
volume = {28}, 
note = {Publisher: American Institute of Physics doi: 10.1063/1.1744102}, 
month = {02}, 
}

@article{flory_1942, 
year = {1942},  
title = {{Thermodynamics of High Polymer Solutions}}, 
author = {Flory, Paul J}, 
journal = {J. Chem. Phys.}, 
issn = {0021-9606}, 
doi = {10.1063/1.1723621}, 
url = {https://doi.org/10.1063/1.1723621}, 
pages = {51--61}, 
number = {1}, 
volume = {10}, 
month = {01}, 
}

@article{huggins_1941, 
year = {1941}, 
title = {{Solutions of Long Chain Compounds}}, 
author = {Huggins, Maurice L}, 
journal = {J. Chem. Phys.}, 
issn = {0021-9606}, 
doi = {10.1063/1.1750930}, 
url = {https://doi.org/10.1063/1.1750930}, 
pages = {440--440}, 
number = {5}, 
volume = {9}, 
month = {05}, 
}

@article{flory_1941, 
year = {1941}, 
title = {{Thermodynamics of High Polymer Solutions}}, 
author = {Flory, Paul J}, 
journal = {J. Chem. Phys.}, 
issn = {0021-9606}, 
doi = {10.1063/1.1750971}, 
url = {https://doi.org/10.1063/1.1750971}, 
pages = {660--660}, 
number = {8}, 
volume = {9}, 
}

@article{taylor_1934, 
title={The formation of emulsions in definable fields of flow}, 
volume={146}, 
ISSN={0950-1207}, 
url={https://doi.org/10.1098/rspa.1934.0169}, 
DOI={10.1098/rspa.1934.0169}, 
number={858}, 
journal={Proc. R. Soc. Lond. A}, 
author={Taylor, Geoffrey Ingram}, 
year={1934}, 
month=jan, 
pages={501–523} 
}

@article{bray_1994, 
title={Theory of phase-ordering kinetics}, 
volume={43}, 
ISSN={0001-8732}, 
url={https://doi.org/10.1080/00018739400101505}, 
DOI={10.1080/00018739400101505}, 
number={3}, 
journal={Advances in Physics}, 
author={Bray, A.J.}, 
year={1994}, 
pages={357–459} 
}

@article{stone_1994, 
title={Dynamics of Drop Deformation and Breakup in Viscous Fluids}, 
volume={26}, 
ISSN={0066-4189}, 
url={https://www.annualreviews.org/content/journals/10.1146/annurev.fl.26.010194.000433}, 
DOI={10.1146/annurev.fl.26.010194.000433}, 
number={1}, 
journal={Annu. Rev. Fluid Mech.}, 
author={Stone, Howard A}, 
year={1994}, 
pages={65–102} 
}

@article{adkins_2022, 
title={Dynamics of active liquid interfaces}, 
volume={377}, 
ISSN={0036-8075}, 
url={https://doi.org/10.1126/science.abo5423}, 
DOI={10.1126/science.abo5423}, 
number={6607}, 
journal={Science}, 
author={Adkins, Raymond and Kolvin, Itamar and You, Zhihong and Witthaus, Sven and Marchetti, M. Cristina and Dogic, Zvonimir}, 
year={2022}, 
month=aug, 
pages={768–772} 
}

@article{caballero_2022, 
title={Activity-Suppressed Phase Separation}, 
volume={129}, 
ISSN={0031-9007}, 
url={https://link.aps.org/doi/10.1103/PhysRevLett.129.268002},
DOI={10.1103/physrevlett.129.268002}, 
number={26}, 
journal={Physical Review Letters}, 
author={Caballero, Fernando and Marchetti, M. Cristina}, 
year={2022}, 
month=dec, 
pages={268002} 
}

@article{strom_2024, 
title={Condensate interfacial forces reposition DNA loci and probe chromatin viscoelasticity},
volume={187},
ISSN={0092-8674}, 
url={https://www.sciencedirect.com/science/article/pii/S0092867424008286}, 
DOI={10.1016/j.cell.2024.07.034}, 
number={19}, 
journal={Cell}, 
author={Strom, Amy R. and Kim, Yoonji and Zhao, Hongbo and Chang, Yi-Che and Orlovsky, Natalia D. and Košmrlj, Andrej and Storm, Cornelis and Brangwynne, Clifford P.}, 
year={2024}, 
month=sep, 
pages={5282-5297.e20} 
}

@article{fletcher_2010, 
title={Cell mechanics and the cytoskeleton}, 
volume={463}, 
ISSN={1476-4687}, 
url={https://doi.org/10.1038/nature08908}, 
DOI={10.1038/nature08908}, 
number={7280}, 
journal={Nature}, 
author={Fletcher, Daniel A. and Mullins, R. Dyche}, 
year={2010}, 
month=jan, 
pages={485–492} 
}

@article{robinson_2024, 
title={Universal limiting behaviour of reaction-diffusion systems with conservation laws}, 
DOI={10.48550/arxiv.2406.02409}, 
journal={arXiv}, 
author={Robinson, Joshua F and Machon, Thomas and Speck, Thomas}, 
year={2024} 
}

@article{Bressloff_2020, 
title={Two-dimensional droplet ripening in a concentration gradient}, 
volume={53}, 
ISSN={1751-8113}, 
url={https://dx.doi.org/10.1088/1751-8121/aba39a}, 
DOI={10.1088/1751-8121/aba39a}, 
number={36}, 
journal={J. Phys. A}, 
author={Bressloff, Paul C}, 
year={2020}, 
month=aug, 
pages={365002} 
}

@article{kumar_2023,
title = {Fluctuations and Shape Dependence of Microphase Separation in Systems with Long-Range Interactions},
author = {Kumar, Amit and Safran, Samuel A.},
journal = {Phys. Rev. Lett.},
volume = {131},
issue = {25},
pages = {258401},
numpages = {7},
year = {2023},
month = {Dec},
publisher = {American Physical Society},
doi = {10.1103/PhysRevLett.131.258401},
url = {https://link.aps.org/doi/10.1103/PhysRevLett.131.258401}
}

@article{zwicker_2025, 
title={Physics of droplet regulation in biological cells}, 
volume={88}, 
ISSN={0034-4885}, 
DOI={10.1088/1361-6633/ae12a7}, 
number={11}, 
journal={Rep. Prog. Phys.}, 
author={Zwicker, David and Paulin, Oliver W and Burg, Cathelijne ter}, 
year={2025}, 
pages={116601} 
}

@article{decayeux_2021, 
title={Spontaneous propulsion of an isotropic colloid in a phase-separating environment}, 
volume={104}, 
ISSN={2470-0045}, 
url={https://link.aps.org/doi/10.1103/PhysRevE.104.034602}, 
DOI={10.1103/physreve.104.034602}, 
number={3}, 
journal={Physical Review E}, 
author={Decayeux, Jeanne and Dahirel, Vincent and Jardat, Marie and Illien, Pierre}, 
year={2021}, 
month=sep, 
pages={034602} 
}

@article{demarchi_2023,
title = {Enzyme-Enriched Condensates Show Self-Propulsion, Positioning, and Coexistence},
author = {Demarchi, Leonardo and Goychuk, Andriy and Maryshev, Ivan and Frey, Erwin},
journal = {Phys. Rev. Lett.},
volume = {130},
issue = {12},
pages = {128401},
numpages = {6},
year = {2023},
month = {Mar},
publisher = {American Physical Society},
doi = {10.1103/PhysRevLett.130.128401},
url = {https://doi.org/10.1103/PhysRevLett.130.128401},
}

@Article{li_2020,
author    = {Li, Yuting I. and Cates, Michael E.},
journal   = {J. Stat. Mech.},
title     = {Non-equilibrium phase separation with reactions: a canonical model and its behaviour},
year      = {2020},
month     = may,
number    = {5},
pages     = {053206},
volume    = {2020},
doi       = {10.1088/1742-5468/ab7e2d},
url       = {https://doi.org/10.1088/1742-5468/ab7e2d},
publisher = {{IOP} Publishing},
}

@article{weber_2017, 
title={Droplet ripening in concentration gradients}, 
volume={19}, 
DOI={10.1088/1367-2630/aa6b84}, 
url={https://doi.org/10.1088/1367-2630/aa6b84},
number={5}, 
journal={New J. Phys.}, 
author={Weber, Christoph A and Lee, Chiu Fan and Jülicher, Frank}, 
year={2017}, 
pages={053021} 
}

@article{bruinsma_2014, 
title={Chromatin Hydrodynamics}, 
volume={106}, 
ISSN={0006-3495}, 
url={https://doi.org/10.1016/j.bpj.2014.03.038}, 
DOI={10.1016/j.bpj.2014.03.038}, 
number={9}, 
journal={Biophys. J.}, 
author={Bruinsma, Robijn and Grosberg, Alexander Y. and Rabin, Yitzhak and Zidovska, Alexandra}, 
year={2014}, 
pages={1871–1881} 
}

@article{shaban_2018, 
title={Formation of correlated chromatin domains at nanoscale dynamic resolution during transcription}, 
volume={46}, 
ISSN={0305-1048}, 
url={https://doi.org/10.1093/nar/gky269}, 
DOI={10.1093/nar/gky269}, 
number={13}, 
journal={Nucleic Acids Res.}, 
author={Shaban, Haitham A. and Barth, Roman and Bystricky, Kerstin}, 
year={2018}, 
pages={e77–e77} 
}

@article{zidovska_2013, 
title={Micron-scale coherence in interphase chromatin dynamics}, 
volume={110}, 
url={https://doi.org/10.1073/pnas.1220313110}, 
DOI={10.1073/pnas.1220313110},  
number={39}, 
journal={Proc. Natl. Acad. Sci. U.S.A.}, 
author={Zidovska, Alexandra and Weitz, David A. and Mitchison, Timothy J.}, 
year={2013}, 
month=sep, 
pages={15555–15560}
}

@article{yu_2023, 
title={Pattern formation of lipid domains in bilayer membranes}, 
volume={21}, 
ISSN={1744-683X}, 
url={http://dx.doi.org/10.1039/D5SM00276A}, 
DOI={10.1039/d5sm00276a}, 
number={21}, 
journal={Soft Matter}, 
author={Yu, Qiwei and Košmrlj, Andrej}, 
year={2025}, 
pages={4288–4297} 
}

@article{yuan_2021, 
title={Membrane bending by protein phase separation}, 
volume={118}, 
ISSN={0027-8424}, 
url={https://doi.org/10.1073/pnas.2017435118},
DOI={10.1073/pnas.2017435118}, 
number={11}, 
journal={Proc. Natl. Acad. Sci. U.S.A.}, 
author={Yuan, Feng and Alimohamadi, Haleh and Bakka, Brandon and Trementozzi, Andrea N. and Day, Kasey J. and Fawzi, Nicolas L. and Rangamani, Padmini and Stachowiak, Jeanne C.}, 
year={2021}, 
month=mar, 
pages={e2017435118} 
}

@article{style_2018, 
title={Liquid-Liquid Phase Separation in an Elastic Network}, 
volume={8}, 
url={https://link.aps.org/doi/10.1103/PhysRevX.8.011028},
DOI={10.1103/physrevx.8.011028}, 
number={1}, 
journal={Phys. Rev. X}, 
author={Style, Robert W. and Sai, Tianqi and Fanelli, Nicoló and Ijavi, Mahdiye and Smith-Mannschott, Katrina and Xu, Qin and Wilen, Lawrence A. and Dufresne, Eric R.}, 
year={2018}, 
month=feb, 
pages={011028} 
}

@article{brangwynne_2009_active_diffusion, 
title={Intracellular transport by active diffusion}, 
volume={19}, 
ISSN={0962-8924}, 
url={https://doi.org/10.1016/j.tcb.2009.04.004}, 
DOI={10.1016/j.tcb.2009.04.004}, 
number={9}, 
journal={Trends Cell Biol.}, 
author={Brangwynne, Clifford P. and Koenderink, Gijsje H. and MacKintosh, Frederick C. and Weitz, David A.}, 
year={2009}, 
pages={423–427} 
}

@article{winter_2024, 
title={Phase separation on deformable membranes: Interplay of mechanical coupling and dynamic surface geometry}, 
volume={111}, 
ISSN={2470-0045}, 
url={https://link.aps.org/doi/10.1103/PhysRevE.111.044405}, 
DOI={10.1103/physreve.111.044405}, 
number={4}, 
journal={Phys. Rev. E}, 
author={Winter, Antonia and Liu, Yuhao and Ziepke, Alexander and Dadunashvili, George and Frey, Erwin}, 
year={2025}, 
month=apr, 
pages={044405} 
}

@article{qiang_2024, 
title={Nonlocal Elasticity Yields Equilibrium Patterns in Phase Separating Systems}, 
volume={14}, 
url={https://link.aps.org/doi/10.1103/PhysRevX.14.021009}, 
DOI={10.1103/physrevx.14.021009}, 
number={2}, 
journal={Phys. Rev. X}, 
author={Qiang, Yicheng and Luo, Chengjie and Zwicker, David}, 
year={2024}, 
month=apr, 
pages={021009} 
}

@article{rosowski_2020, 
title={Elastic ripening and inhibition of liquid–liquid phase separation}, 
volume={16},
ISSN={1745-2473}, 
url={https://doi.org/10.1038/s41567-019-0767-2}, 
DOI={10.1038/s41567-019-0767-2}, 
number={4}, 
journal={Nat. Phys.}, 
author={Rosowski, Kathryn A. and Sai, Tianqi and Vidal-Henriquez, Estefania and Zwicker, David and Style, Robert W. and Dufresne, Eric R.}, 
year={2020}, 
month=apr, 
pages={422–425} 
}

@article{quail_2021, 
title={Force generation by protein–DNA co-condensation}, 
volume={17}, 
ISSN={1745-2473}, 
url={https://doi.org/10.1038/s41567-021-01285-1}, 
DOI={10.1038/s41567-021-01285-1}, 
number={9}, 
journal={Nat. Phys.}, 
author={Quail, Thomas and Golfier, Stefan and Elsner, Maria and Ishihara, Keisuke and Murugesan, Vasanthanarayan and Renger, Roman and Jülicher, Frank and Brugués, Jan}, 
year={2021}, 
month=sep, 
pages={1007–1012} 
}

@article{graham_2024, 
title={Liquid-like condensates mediate competition between actin branching and bundling}, 
volume={121},
ISSN={0027-8424}, 
url={https://doi.org/10.1073/pnas.2309152121}, 
DOI={10.1073/pnas.2309152121}, 
number={3}, 
journal={Proc. Natl. Acad. Sci. U.S.A.}, 
author={Graham, Kristin and Chandrasekaran, Aravind and Wang, Liping and Yang, Noel and Lafer, Eileen M. and Rangamani, Padmini and Stachowiak, Jeanne C.}, 
year={2024}, 
month=jan, 
pages={e2309152121} 
}

@article{jambonpuillet_2023, 
title={Phase-separated droplets swim to their dissolution}, 
volume={15}, 
ISSN={2041-1723}, 
DOI={10.1038/s41467-024-47889-y}, 
number={1}, 
journal={Nat. Commun.}, 
author={Jambon-Puillet, Etienne and Testa, Andrea and Lorenz, Charlotta and Style, Robert W. and Rebane, Aleksander A. and Dufresne, Eric R.}, 
year={2024}, 
month=may, 
pages={3919} 
}

@article{wiegand_2020, 
title={Drops and fibers — how biomolecular condensates and cytoskeletal filaments influence each other}, 
volume={4}, 
ISSN={2397-8554}, 
url={https://doi.org/10.1042/ETLS20190174}, 
DOI={10.1042/etls20190174}, 
number={3}, 
journal={Emerg. Top. Life Sci.}, 
author={Wiegand, Tina and Hyman, Anthony A.}, 
year={2020}, 
pages={247–261} 
}

@article{mohapatra_2023, 
title={Biomolecular condensation involving the cytoskeleton}, 
volume={194},
ISSN={0361-9230}, 
url={https://www.sciencedirect.com/science/article/pii/S0361923023000114}, 
DOI={10.1016/j.brainresbull.2023.01.009}, 
journal={Brain Res. Bull.}, 
author={Mohapatra, Satabdee and Wegmann, Susanne}, 
year={2023},
month=mar, 
pages={105–117} 
}

@article {banani_2024,
author = {Banani, Salman F. and Goychuk, Andriy and Natarajan, Pradeep and Zheng, Ming M. and Dall{\textquoteright}Agnese, Giuseppe and Henninger, Jonathan E. and Kardar, Mehran and Young, Richard A. and Chakraborty, Arup K.},
title = {Active {RNA} synthesis patterns nuclear condensates},
elocation-id = {2024.10.12.614958},
year = {2024},
doi = {10.1101/2024.10.12.614958},
publisher = {Cold Spring Harbor Laboratory},
URL = {https://www.biorxiv.org/content/early/2024/10/13/2024.10.12.614958},
journal = {bioRxiv}
}

@article{lafontaine_2021, 
title={The nucleolus as a multiphase liquid condensate}, 
volume={22}, 
ISSN={1471-0080}, 
DOI={10.1038/s41580-020-0272-6}, 
number={3}, 
journal={Nat. Rev. Mol. Cell Biol.}, 
author={Lafontaine, Denis L. J. and Riback, Joshua A. and Bascetin, Rümeyza and Brangwynne, Clifford P.}, 
year={2021}, 
month={Mar}, 
pages={165–182} 
}

@article{hirose_2023, 
title={A guide to membraneless organelles and their various roles in gene regulation}, 
volume={24}, 
ISSN={1471-0072}, 
url={https://doi.org/10.1038/s41580-022-00558-8}, 
DOI={10.1038/s41580-022-00558-8}, 
number={4}, 
journal={Nat. Rev. Mol. Cell Biol.},
author={Hirose, Tetsuro and Ninomiya, Kensuke and Nakagawa, Shinichi and Yamazaki, Tomohiro},
year={2023}, 
month=apr, 
pages={288–304} 
}

@article{banani_2017, 
title={Biomolecular condensates: organizers of cellular biochemistry}, 
volume={18},
ISSN={1471-0072}, 
url={https://doi.org/10.1038/nrm.2017.7},
DOI={10.1038/nrm.2017.7}, 
number={5}, 
journal={Nat. Rev. Mol. Cell Biol.}, 
author={Banani, Salman F. and Lee, Hyun O. and Hyman, Anthony A. and Rosen, Michael K.}, 
year={2017}, 
month=may, 
pages={285–298} 
}

@article{choi_2020, 
title={Physical Principles Underlying the Complex Biology of Intracellular Phase Transitions}, 
volume={49}, 
ISSN={1936-122X}, 
url={https://www.annualreviews.org/content/journals/10.1146/annurev-biophys-121219-081629},
DOI={10.1146/annurev-biophys-121219-081629}, 
number={1}, 
journal={Annu. Rev. Biophys.}, 
author={Choi, Jeong-Mo and Holehouse, Alex S. and Pappu, Rohit V.},
year={2020}, 
pages={1–27} 
}

@article{alberti_2019, 
title={Considerations and Challenges in Studying Liquid-Liquid Phase Separation and Biomolecular Condensates}, 
volume={176}, 
ISSN={0092-8674}, 
url={https://www.sciencedirect.com/science/article/pii/S0092867418316490}, 
DOI={10.1016/j.cell.2018.12.035}, 
number={3}, 
journal={Cell}, 
author={Alberti, Simon and Gladfelter, Amy and Mittag, Tanja}, 
year={2019}, 
month=jan,
pages={419–434} 
}

@article{hyman_2014, 
title={Liquid-Liquid Phase Separation in Biology}, 
volume={30}, 
ISSN={1081-0706},
url={https://www.annualreviews.org/content/journals/10.1146/annurev-cellbio-100913-013325}, 
DOI={10.1146/annurev-cellbio-100913-013325}, 
number={1}, 
journal={Annu. Rev. Cell Dev. Biol.}, 
author={Hyman, Anthony A. and Weber, Christoph A. and Jülicher, Frank}, 
year={2014},
pages={39–58} 
}

@article{shin_2017, 
title={Liquid phase condensation in cell physiology and disease}, 
volume={357}, 
ISSN={0036-8075}, 
url={https://doi.org/10.1126/science.aaf4382},
DOI={10.1126/science.aaf4382},
number={6357}, 
journal={Science}, 
author={Shin, Yongdae and Brangwynne, Clifford P.}, 
year={2017}, 
month=sep, 
pages={eaaf4382} 
}

@article{brangwynne_2009, 
title={Germline {P} Granules Are Liquid Droplets That Localize by Controlled Dissolution/Condensation}, 
volume={324}, 
ISSN={0036-8075},
url={https://doi.org/10.1126/science.1172046}, 
DOI={10.1126/science.1172046}, 
number={5935}, 
journal={Science}, 
author={Brangwynne, Clifford P. and Eckmann, Christian R. and Courson, David S. and Rybarska, Agata and Hoege, Carsten and Gharakhani, Jöbin and Jülicher, Frank and Hyman, Anthony A.}, 
year={2009}, 
month=jun, 
pages={1729–1732} 
}

@article{lyon_2020, 
title={A framework for understanding the functions of biomolecular condensates across scales}, 
volume={22}, 
ISSN={1471-0072}, 
url={https://doi.org/10.1038/s41580-020-00303-z}, 
DOI={10.1038/s41580-020-00303-z}, 
number={3}, 
journal={Nat. Rev. Mol. Cell Biol.}, 
author={Lyon, Andrew S. and Peeples, William B. and Rosen, Michael K.}, 
year={2021}, 
month=mar, 
pages={215–235} 
}

@article{faber_2022, 
title={Nuclear speckles – a driving force in gene expression}, 
volume={135}, 
ISSN={0021-9533}, 
DOI={10.1242/jcs.259594}, 
number={13}, 
journal={J. Cell Sci.}, 
author={Faber, Gabriel P. and Nadav-Eliyahu, Shani and Shav-Tal, Yaron}, 
year={2022}, 
month=jul, 
pages={jcs259594} 
}

@article{henninger_2021, 
title={{RNA}-Mediated Feedback Control of Transcriptional Condensates},
volume={184}, 
ISSN={0092-8674}, 
DOI={10.1016/j.cell.2020.11.030},  
url = {https://doi.org/10.1016/j.cell.2020.11.030},
number={1}, 
journal={Cell}, 
author={Henninger, Jonathan E. and Oksuz, Ozgur and Shrinivas, Krishna and Sagi, Ido and LeRoy, Gary and Zheng, Ming M. and Andrews, J. Owen and Zamudio, Alicia V. and Lazaris, Charalampos and Hannett, Nancy M. and Lee, Tong Ihn and Sharp, Phillip A. and Cissé, Ibrahim I. and Chakraborty, Arup K. and Young, Richard A.}, 
year={2021}, 
month={Jan}, 
pages={207-225.e24} 
}

@article{shrinivas_2019, 
title={Enhancer Features that Drive Formation of Transcriptional Condensates}, 
volume={75}, 
ISSN={1097-2765}, 
DOI={10.1016/j.molcel.2019.07.009}, 
url = {https://doi.org/10.1016/j.molcel.2019.07.009},
number={3}, 
journal={Mol. Cell}, 
author={Shrinivas, Krishna and Sabari, Benjamin R. and Coffey, Eliot L. and Klein, Isaac A. and Boija, Ann and Zamudio, Alicia V. and Schuijers, Jurian and Hannett, Nancy M. and Sharp, Phillip A. and Young, Richard A. and Chakraborty, Arup K.}, 
year={2019}, 
month={Aug}, 
pages={549-561.e7} 
}

@article{hnisz_2017, 
title={A Phase Separation Model for Transcriptional Control}, 
volume={169}, 
ISSN={0092-8674}, 
DOI={10.1016/j.cell.2017.02.007}, 
number={1}, 
journal={Cell}, 
author={Hnisz, Denes and Shrinivas, Krishna and Young, Richard A. and Chakraborty, Arup K. and Sharp, Phillip A.}, 
year={2017}, 
month=mar, 
pages={13–23} 
}

@article{sabari_2018, 
title={Coactivator condensation at super-enhancers links phase separation and gene control}, 
volume={361}, 
DOI={doi:10.1126/science.aar3958}, 
number={6400}, 
journal={Science}, 
author={Sabari, Benjamin R. and Dall’Agnese, Alessandra and Ann Boija, Ann and Klein, Isaac A. and Coffey, Eliot L. and Shrinivas, Krishna and Abraham, Brian J. and Hannett, Nancy M. and Zamudio, Alicia V. and Manteiga, John C. and Li, Charles H. and Guo, Yang E. and Day, Daniel S. and Schuijers, Jurian and Vasile, Eliza and Malik, Sohail and Hnisz, Denes and Lee, Tong Ihn and Cisse, Ibrahim I. and Roeder, Robert G. and Sharp, Phillip A. and Chakraborty, Arup K. and Young, Richard A.}, 
year={2018}, 
pages={eaar3958} 
}

@article{Meng_2018, 
title={Clustering of Magnetic Swimmers in a Poiseuille Flow}, 
volume={120}, 
ISSN={0031-9007}, 
url={https://link.aps.org/doi/10.1103/PhysRevLett.120.188101},
DOI={10.1103/physrevlett.120.188101}, 
number={18}, 
journal={Phys. Rev. Lett.}, 
author={Meng, Fanlong and Matsunaga, Daiki and Golestanian, Ramin}, 
year={2018}, 
month=may, 
pages={188101} 
}

@article{ro_2021, 
title={Disorder-Induced Long-Ranged Correlations in Scalar Active Matter}, 
volume={126}, 
ISSN={0031-9007}, 
url={https://link.aps.org/doi/10.1103/PhysRevLett.126.048003},
DOI={10.1103/physrevlett.126.048003}, 
number={4}, 
journal={Phys. Rev. Lett.}, 
author={Ro, Sunghan and Kafri, Yariv and Kardar, Mehran and Tailleur, Julien}, 
year={2021}, 
month=jan, 
pages={048003} 
}

@article{baek_2018, 
title={Generic Long-Range Interactions Between Passive Bodies in an Active Fluid}, 
volume={120}, 
url={https://link.aps.org/doi/10.1103/PhysRevLett.120.058002},
DOI={10.1103/physrevlett.120.058002}, 
number={5}, 
journal={Phys. Rev. Lett.}, 
author={Baek, Yongjoo and Solon, Alexandre P. and Xu, Xinpeng and Nikola, Nikolai and Kafri, Yariv},
year={2018}, 
month=jan, 
pages={058002} 
}

@article{schwarz_2002, 
title={Elastic Interactions of Cells}, 
volume={88}, 
ISSN={0031-9007}, 
url={https://link.aps.org/doi/10.1103/PhysRevLett.88.048102}, DOI={10.1103/physrevlett.88.048102}, 
number={4}, 
journal={Phys. Rev. Lett.}, 
author={Schwarz, U. S. and Safran, S. A.}, 
year={2002}, 
month=jan, 
pages={048102} 
}

@article{bose_2022, 
title={Collective States of Active Particles With Elastic Dipolar Interactions}, 
volume={10}, 
DOI={10.3389/fphy.2022.876126}, 
journal={Front. Phys.}, 
author={Bose, Subhaya and Noerr, Patrick S. and Gopinathan, Ajay and Gopinath, Arvind and Dasbiswas, Kinjal}, 
year={2022}, 
pages={876126} 
}

@article{doan_2024, 
title={Diffusiophoresis promotes phase separation and transport of biomolecular condensates}, 
volume={15}, 
ISSN={2041-1723}, 
url={https://doi.org/10.1038/s41467-024-51840-6}, 
DOI={10.1038/s41467-024-51840-6}, 
number={1}, 
journal={Nat. Commun.}, 
author={Doan, Viet Sang and Alshareedah, Ibraheem and Singh, Anurag and Banerjee, Priya R. and Shin, Sangwoo}, 
year={2024}, 
month=sep, 
pages={7686} 
}

@article{zwanzig_1964_fluctuating_hydro,
title={Hydrodynamic Fluctuations and Stokes' Law Friction},
author={Robert W. Zwanzig},
journal={J. Res. Natl. Bur. Stand. B},
year={1964},
pages={143},
url={https://doi.org/10.6028/jres.068b.019},
doi={10.6028/jres.068b.019}
}

@article{aljord_2022, 
title={Cytoplasmic forces functionally reorganize nuclear condensates in oocytes}, 
volume={13},
ISSN={2041-1723}, 
DOI={10.1038/s41467-022-32675-5}, 
number={1}, 
journal={Nat. Commun.}, 
author={Al Jord, Adel and Letort, Ga\"elle and Chanet, Soline and Tsai, Feng-Ching and Antoniewski, Christophe and Eichmuller, Adrien and Da Silva, Christelle and Huynh, Jean-René and Gov, Nir S. and Voituriez, Rapha\"el and Terret, Marie-\'Emilie and Verlhac, Marie-H\'el\`ene}, 
year={2022}, 
month=aug, 
pages={5070} 
}

@article{rotne_prager_1969, 
title={Variational Treatment of Hydrodynamic Interaction in Polymers}, 
volume={50}, 
ISSN={0021-9606}, 
url={https://doi.org/10.1063/1.1670977}, 
DOI={10.1063/1.1670977}, 
number={11}, 
journal={J. Chem. Phys.}, 
author={Rotne, Jens and Prager, Stephen}, 
year={1969}, 
pages={4831–4837} 
}

@article{yamakawa_1970, 
title={Transport Properties of Polymer Chains in Dilute Solution: Hydrodynamic Interaction}, 
volume={53}, 
ISSN={0021-9606}, 
url={https://doi.org/10.1063/1.1673799}, 
DOI={10.1063/1.1673799}, 
number={1}, 
journal={J. Chem. Phys.}, 
author={Yamakawa, Hiromi}, 
year={1970}, 
pages={436–443} 
}

@article{goychuk_2024_self_consistent,
title = {Self-consistent sharp interface theory of active condensate dynamics},
author = {Goychuk, Andriy and Demarchi, Leonardo and Maryshev, Ivan and Frey, Erwin},
journal = {Phys. Rev. Res.},
volume = {6},
issue = {3},
pages = {033082},
numpages = {25},
year = {2024},
month = {Jul},
publisher = {American Physical Society},
doi = {10.1103/PhysRevResearch.6.033082},
url = {https://link.aps.org/doi/10.1103/PhysRevResearch.6.033082}
}

@article{goh_2024, 
title={{RNA} gradients can guide condensates toward promoters: Implications for enhancer–promoter contacts and condensate-promoter kissing}, 
volume={163}, 
ISSN={0021-9606}, 
DOI={10.1063/5.0277838}, 
number={10}, 
journal={J. Chem. Phys.}, 
author={Goh, David and Kannan, Deepti and Natarajan, Pradeep and Goychuk, Andriy and Chakraborty, Arup K.}, 
year={2025}, 
pages={104905} 
}

@article{hohenberg_halperin_1977,
title = {Theory of dynamic critical phenomena},
author = {Hohenberg, P. C. and Halperin, B. I.},
journal = {Rev. Mod. Phys.},
volume = {49},
issue = {3},
pages = {435--479},
numpages = {0},
year = {1977},
month = {Jul},
publisher = {American Physical Society},
doi = {10.1103/RevModPhys.49.435},
url = {https://link.aps.org/doi/10.1103/RevModPhys.49.435}
}

@article{brinkman_1949, 
title={A calculation of the viscous force exerted by a flowing fluid on a dense swarm of particles}, 
volume={1}, 
ISSN={1573-1987}, 
DOI={10.1007/BF02120313}, 
number={1}, 
journal={Flow Turbul. Combust.}, 
author={Brinkman, H. C.}, 
year={1949}, 
month=dec, 
pages={27–34} 
}

@article{zhang_2024_brownian,
title = {Brownian motion of droplets induced by thermal noise},
author = {Zhang, Haodong and Wang, Fei and Ratke, Lorenz and Nestler, Britta},
journal = {Phys. Rev. E},
volume = {109},
issue = {2},
pages = {024208},
numpages = {14},
year = {2024},
month = {Feb},
publisher = {American Physical Society},
doi = {10.1103/PhysRevE.109.024208},
url = {https://link.aps.org/doi/10.1103/PhysRevE.109.024208}
}

@article{Kubo_1966,
doi = {10.1088/0034-4885/29/1/306},
url = {https://dx.doi.org/10.1088/0034-4885/29/1/306},
year = {1966},
month = {jan},
publisher = {Institute of Physics},
volume = {29},
number = {1},
pages = {255},
author = {R Kubo},
title = {The fluctuation-dissipation theorem},
journal = {Rep. Prog. Phys.}
}

\end{document}